\documentclass[preprint,12pt,a4paper]{elsarticle}

\usepackage{amssymb}
\usepackage{amsmath}
\usepackage{epstopdf}
\usepackage{float}
\usepackage{algorithm}
\usepackage{algpseudocode}
\usepackage{amsfonts}
\usepackage{numprint}
\usepackage{wrapfig}
\usepackage{a4wide}

\npdecimalsign{.}
\npthousandsep{}
\nprounddigits{4}

\journal{PhD thesis}
\date{May 31, 2021}

\newcommand*\diff{\mathop{}\!\mathrm{d}}
\newcommand{\mbu}{\mathbf{u}}
\newcommand{\nab}{\nabla}
\newcommand{\mbf}{\mathbf{f}}
\newcommand{\mbV}{\mathbf{V}}
\newcommand{\mbv}{\mathbf{v}}
\newcommand{\mbn}{\mathbf{n}}
\newcommand{\mbx}{\mathbf{x}}

\begin{document}

\begin{frontmatter}



\title{Multi-mesh multi-objective optimization \\ with application to a model problem in urban design}


\author[chalmers]{Anders Logg} \ead{logg@chalmers.se}
\author[chalmers]{Christian Valdemar Lorenzen\fnref{fn1}} \ead{christianvaldemar@icloud.com}
\author[chalmers]{Carl Lundholm\corref{cor1}} \ead{carlun@chalmers.se}

\fntext[fn1]{Family name: Lorenzen. Was visiting, original affiliation: \emph{Department of Mathematics and Computer Science, University of Southern Denmark, DK-5230 Odense M, Denmark}}

\cortext[cor1]{\emph{Corresponding author} (Telephone: +46 (0)31 772 53 62, Fax: +46 (0)31-16 19 73)}

\address[chalmers]{Department of Mathematical Sciences, \\
                   Chalmers University of Technology and University of Gothenburg, \\
                   SE-412 96 Gothenburg, Sweden}


\begin{abstract}
We present an application of multi-mesh finite element methods as part of
a methodology for optimizing settlement layouts. By formulating a multi-objective optimization problem, we demonstrate how a given number of buildings may be optimally placed on a given piece of land with respect to both wind conditions and the view experienced from the buildings.
The wind flow is modeled by a multi-mesh (cut finite element) method.
This allows each building to be embedded in a boundary-fitted mesh which can be moved freely on top of a fixed background mesh. This approach enables a
multitude of settlement layouts to be evaluated without the need for
costly mesh generation when changing the configuration of buildings. The view is modeled by a measure that takes into account the totality of unobstructed view from the collection of buildings, and is efficiently computed by rasterization.
\end{abstract}



\end{frontmatter}

\section{Introduction}

When designing settlement layouts, architects need to take a large
number of variables into consideration, such as economic interests,
connections to infrastructure (roads, water and electricity),
experienced quality of view, wind conditions, and
many more, see, e.g.,
\cite{Batty2001, Blocken2012, Hang2012, Shi2015, jaillot2017generic}.
One particular variable to consider is urban wind comfort, which can be evaluated by CFD. There are several examples of urban CFD simulations in the literature, see, e.g., \cite{Baskaran:1996aa,Blocken:2007aa,Blocken:2011aa,
Heuveline:2011aa, Blocken2014, Ingelsten:2016aa}.

A central issue when constructing a computational tool for urban design
is that the tool should be able to quickly evaluate a
multitude of suggested configurations, either as part of an optimization loop
or as part of a manual (artistic) iterative design process. Standard
numerical methods for flow computations require that a mesh or
grid is generated around buildings, ground and other
objects. Generating such a body-fitted mesh is a costly procedure and
even more so when a large number of different meshes must be created,
one for each configuration of the buildings. A way to reduce the cost of
mesh generation is to use \emph{multi-mesh} finite element
methods, see, e.g., \cite{Hansbo:2003aa, Massing:2014aa,
Johansson2018}. multi-mesh finite element methods allow a problem to
be posed on a collection of meshes that may overlap arbitrarily and which together define the
computational domain, see \cite{Dokken} for an interesting application in shape optimization.
The overlapping meshes may be moved freely relative
to one another, allowing a multitude of configurations without costly mesh generation.
However, this flexibility comes at a price.
If the finite element method is not carefully designed, certain
configurations may lead to very ill-conditioned systems, low accuracy
and even blow-up. Another concern is that these methods require
integration over cut cells and interfaces, resulting in challenging
computational geometry problems, when intersections and
quadrature points must be computed efficiently and robustly.

In this study, we examine how multi-mesh finite element methods may be
applied in the process of designing settlement layouts. The idea is to
use such methods to efficiently compute wind patterns around buildings.
We here employ the recently published multi-mesh finite element formulation for the Stokes equations~\cite{Johansson:2015aa}. The model problem we consider is based on a challenge in urban design presented in \cite{Johansson:2014aa},
where houses are to be placed on the island ``Lilla Fjellsholmen'' which is located off the west coast of Sweden.

In addition to urban wind comfort, we take into account the experienced view
from the buildings. We construct quantitative measures for both 3D and 2D settings that can be used to evaluate the view for any given
design. In addition to standard visibility analysis tools such as isovists
and viewsheds, see, e.g., \cite{Benedikt1979, Yang2007, Nutsford2015, Poerwoningsih2016},
the 3D view measure may not only take into account how much surface
and space that is seen, but also \emph{what} is seen.
This is done by incorporating object weights in the
measure, allowing air, water, ground, buildings, herbage and other
objects to have different influence on the view, an idea also presented by
\cite{Fisher-Gewirtzman2018}. The use of these object weights is also the
reason for using the term `view' instead of, e.g., `visibility'.
The 3D view measure may be computed efficiently by rasterization.
The 2D view measure is a binary version of the 3D measure that does not
use object weights. This makes it equivalent to a simplified isovist.

We combine the wind and view models to solve a multi-objective optimization problem to find optimal building configurations. See \cite{Bruno2010, Keough2010, S.CajiotN.SchulerM.PeterA.Koch2017} for examples of the use of mathematical optimization in urban design and architecture.


In the remainder of this paper, we first present the models for wind
and view, and how to use them in optimization in Section~\ref{sec:methods}.
Results are presented in Section~\ref{sec:results}, and
discussed in relation to the methods in Section~\ref{sec:discussion}.
We present our conclusions, discuss current limitations, and
future work in Section~\ref{sec:conclusions}.

\section{Methods}
\label{sec:methods}

We here present the models used to compute wind and view, and how the
output from these models may be used to optimize settlement layouts.


\subsection{Computation of wind}

To model the wind over the island and houses, a cut finite element method
for Stokes equations on overlapping meshes is used. A physically valid analysis of wind conditions would require the solution of the Navier--Stokes equations but for the current model study, we employ the simpler Stokes equations as a demonstration of the potential of multi-mesh finite element methods in combination with optimization or interactive computing, which both require a multitude of geometric configurations to be solved.

The method is presented in detail by~\cite{Johansson:2015aa}, but to better understand
the application at hand and present necessary terminology, we give a brief summary of
the method here. For a bounded domain $\Omega \subset \mathbb{R}^d$ with boundary
$\partial \Omega$, the strong problem formulation for Stokes equations reads:
Find the velocity $\mbu : \Omega \to \mathbb{R}^d$ and the pressure
$p : \Omega \to \mathbb{R}$ such that
\begin{equation}
\left\{
\begin{split}
- \Delta \mbu + \nab p & = \mbf && \text{in} \ \Omega, \\
\nab \cdot \mbu & = 0 && \text{in} \ \Omega, \\
\mbu & = \mathbf{0} && \text{on} \ \partial\Omega,
\end{split} \label{stokes}
\right.
\end{equation}
\noindent where $\mbf : \Omega \to \mathbb{R}^d$ is a given right-hand
side. To obtain a cut finite element formulation for Stokes equations on
overlapping meshes, we start by considering two bounded domains
$\widehat{\Omega}_0$ and $\widehat{\Omega}_1$, called predomains. Let
$\Omega_0 := \widehat{\Omega}_0 \setminus \widehat{\Omega}_1$ and
$\Omega_1 :=\widehat{\Omega}_1$, with boundaries $\partial \Omega_0$
and $\partial \Omega_1$, respectively. We define the solution domain
by $\Omega := \Omega_0 \cup \Omega_1$ and the joint boundary between
$\Omega_0$ and $\Omega_1$ by
$\Gamma := \partial \Omega_0 \cap \partial \Omega_1$. 
%
%
To obtain a finite element formulation, the predomains,
$\widehat{\Omega}_0$ and $\widehat{\Omega}_1$, are tessellated to
create the meshes $\widehat{\mathcal{K}}_{h,0}$ and
$\widehat{\mathcal{K}}_{h,1}$, respectively, where $h$ denotes mesh size.
We call $\widehat{\mathcal{K}}_{h,0}$ the background mesh, and
$\widehat{\mathcal{K}}_{h,1}$ the overlapping mesh.  
%
%
The following subdomains will also be useful. For $i = 0, 1$,
\begin{align}
\Omega_{h, i} := \bigcup_{\substack{K \in \widehat{\mathcal{K}}_{h,i}, \overline{K} \cap \Omega_i \neq \emptyset}} K, \quad
\omega_{h, 0} := \bigcup_{\substack{K \in \widehat{\mathcal{K}}_{h,0}, \overline{K} \cap \overline{\Omega}_1 = \emptyset}} K
\label{eq_model_meshdomains}
\end{align}
%
%
%
The desired finite element space may then be constructed in three steps:
\begin{enumerate}
\item For $i = 0, 1$, define a product space
$\widehat{\mbV}_{h, i} \times \widehat{Q}_{h, i}$ for each mesh,
$\widehat{\mathcal{K}}_{h,i}$, where $\widehat{\mbV}_{h, i}$ and
$\widehat{Q}_{h, i}$ are finite element spaces for the velocity and
pressure, respectively.
\item Consider the restriction of these spaces to $\Omega_{h,i}$ defined by
\begin{equation}
\mbV_{h, i} \times Q_{h, i} := \widehat{\mbV}_{h, i} |_{\Omega_{h,i}} \times \widehat{Q}_{h, i} |_{\Omega_{h,i}},
\end {equation}
\item The desired space for the finite element formulation is defined by
\begin{equation}
\mbV_h \times Q_h := \bigoplus_{i=0}^1 \mbV_{h, i} \times Q_{h, i}
\label{fespace}
\end{equation}
\end{enumerate}

\noindent Note that functions in $\mbV_h \times Q_h$ will technically have have
two different parts on $\Omega_{h,0} \cap \Omega_1$, nameley one part from
each $\mbV_{h, i} \times Q_{h, i}$, for $i = 0, 1$. However, the convention when
referring to, evaluating, or visualizing functions in $\mbV_h \times Q_h$ is to
always pick the part from the highest mesh in the hierarchy, unless otherwise
specified. Note that this convention allows functions in $\mbV_h \times Q_h$ to
be discontinuous on $\Gamma$ even though the functions in
$\widehat{\mbV}_{h, i} \times \widehat{Q}_{h, i}$, for $i = 0, 1$, are not.
In this application, Taylor-Hood elements are used to ensure the stability of
the solution, i.e., continuous piecewise polynomials of degree two and degree one
for the velocity and the pressure, respectively. The cut finite element formulation
for Stokes equations on two overlapping meshes reads:
Find $(\mbu_h, p_h) \in \mbV_h \times Q_h$ such that
\begin{equation}
\begin{split}
& \quad \, (D\mbu_h, D\mbv)_{\Omega_0} + (D\mbu_h, D\mbv)_{\Omega_1} \\
& - (\langle \partial_\mbn \mbu_h \rangle, [\mbv])_\Gamma - ([\mbu_h], \langle \partial_\mbn \mbv \rangle)_\Gamma \\
& + \beta h^{-1} ([\mbu_h], [\mbv])_\Gamma + ([D\mbu_h], [D\mbv])_{\Omega_{h,0} \cap \Omega_1} \\
& -(\nab \cdot \mbu_h, q)_{\Omega_0} - (\nab \cdot \mbu_h, q)_{\Omega_1} + ([\mbn \cdot \mbu_h], \langle q \rangle)_\Gamma \\
& -(\nab \cdot \mbv, p_h)_{\Omega_0} - (\nab \cdot \mbv, p_h)_{\Omega_1} + ([\mbn \cdot \mbv], \langle p_h \rangle)_\Gamma \\
& + h^2(\Delta \mbu_h - \nab p_h, \Delta \mbv + \nab q)_{\Omega_{h,0}\setminus \omega_{h,0}} \\
= & \quad \, (\mbf, \mbv)_{\Omega_0} + (\mbf, \mbv)_{\Omega_1} - h^2(\mbf, \Delta \mbv + \nab q)_{\Omega_{h,0}\setminus \omega_{h,0}},
\end{split} \label{femstokes}
\end{equation}
\noindent for all $(\mbv, q) \in \mbV_h \times Q_h$. Here, $(\cdot, \cdot)_\Omega$
is the $L^2(\Omega)$-inner product, $D \mbv = (\nab \mbv_1, \dots, \nab \mbv_d)$,
$\langle v \rangle = (v_0 + v_1)/2$ is the average of $v$ on $\Gamma$
($v_i$ is the limit of $v$ on $\Omega_i$ when approaching $\Gamma$, for
$i = 0, 1$),
$\partial_\mbn \mbv = (\nab \mbv_1 \cdot \mbn, \dots, \nab \mbv_d \cdot \mbn)$,
$\mbn$ is the unit normal to $\Gamma$ exterior to $\Omega_1$, $[v] = v_1 - v_0$
is the jump in $v$ on $\Gamma$, $\beta$ is a stability parameter,
$h$ is the mesh size, $\Omega_{h,0} \cap \Omega_1$ is the covered part of all the
background cells that are cut by $\Gamma$, and $\Omega_{h,0} \setminus \omega_{h,0}$
is all the background cells that are cut by $\Gamma$. All
differential operators are to be understood in the generalized sense. For more
details on the method, analysis, and numerical results, see~\cite{Johansson:2015aa}.

For the application studied in this work, the background mesh is also
referred to as the air mesh, since it is a discretization of a region containing air.
The overlapping meshes are also referred to as house meshes,
since they contain the houses.
The multi-mesh finite element model is implemented in Python using the open-source finite element library
FEniCS, see \cite{Logg:2010aa}, whose multi-mesh-functionality provides easy-to-use tools for implementing
cut finite element methods on overlapping meshes. Since multi-mesh formulations are only partly supported in 3D in FEniCS, the current study is limited to the 2D formulation.
We are now ready to formulate the algorithm for obtaining the wind over the
island and houses by solving
(\ref{femstokes}).

\begin{algorithm}[H]
\caption{Wind model}
\label{algoflowmodel}
 \begin{algorithmic}[1]
  \State Geometries for the island and houses are imported. See Figure \ref{figgeos}.
  \State Meshes are generated using the geometries. See Figure \ref{figmeshes}.
  \State The house meshes are placed inside the air mesh. See Figure \ref{figmeshoverlap}.
  \State The linear system of equations, resulting from (\ref{femstokes}), is assembled and
             boundary conditions are applied to it: inlet and outlet for the air mesh, and no-slip for the house
             boundary of the house meshes.
  \State Background cells intersecting house mesh holes are located and marked as covered.
            Corresponding degrees of freedom are inactivated, preventing wind inside houses. See Figure \ref{figautocoverdemo}.
   \State The modified linear system of equations is solved.
 \end{algorithmic}
\end{algorithm}

\begin{figure}[H]
\centering
\includegraphics[width=0.49\linewidth,keepaspectratio]{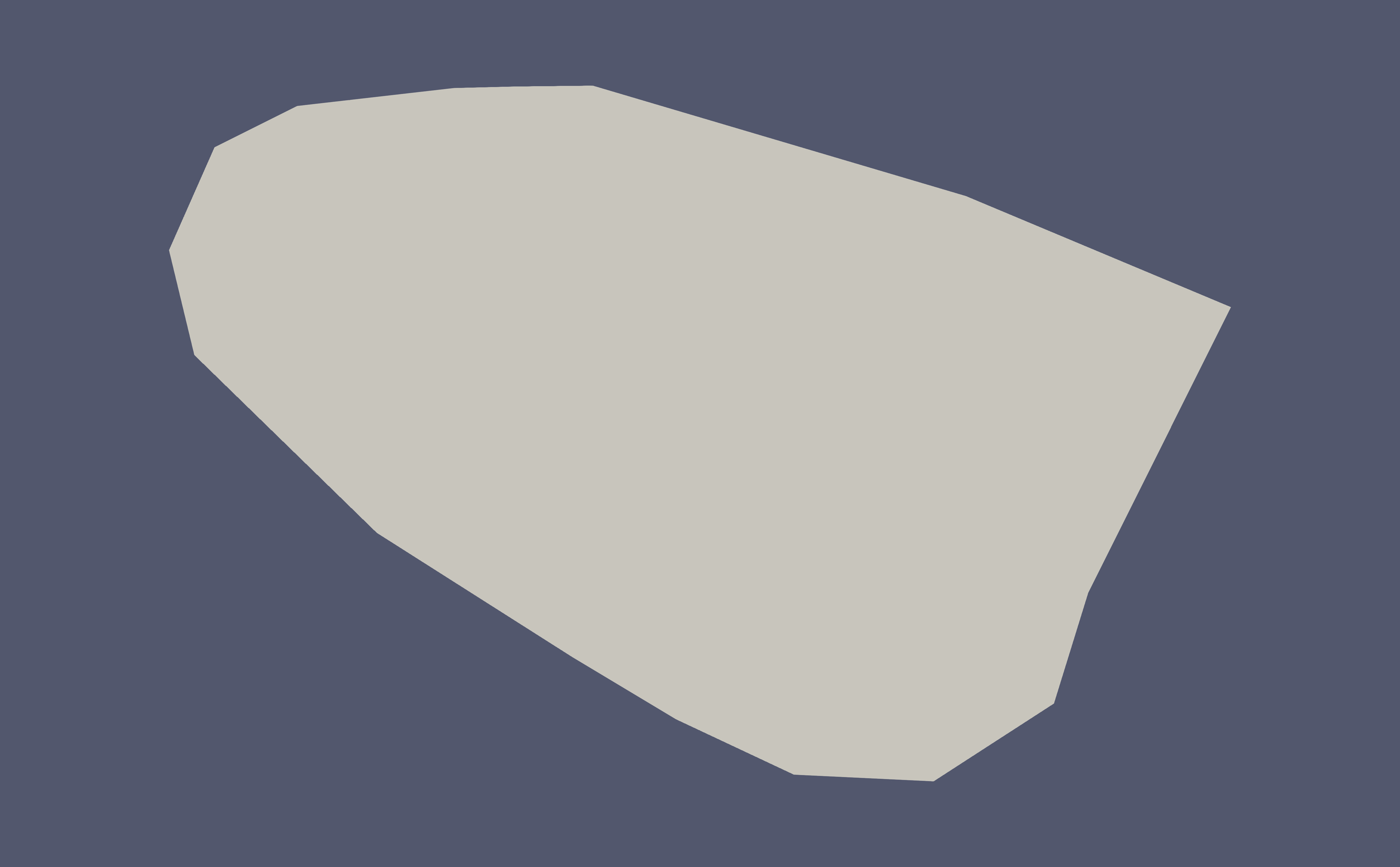}
\includegraphics[width=0.49\linewidth,keepaspectratio]{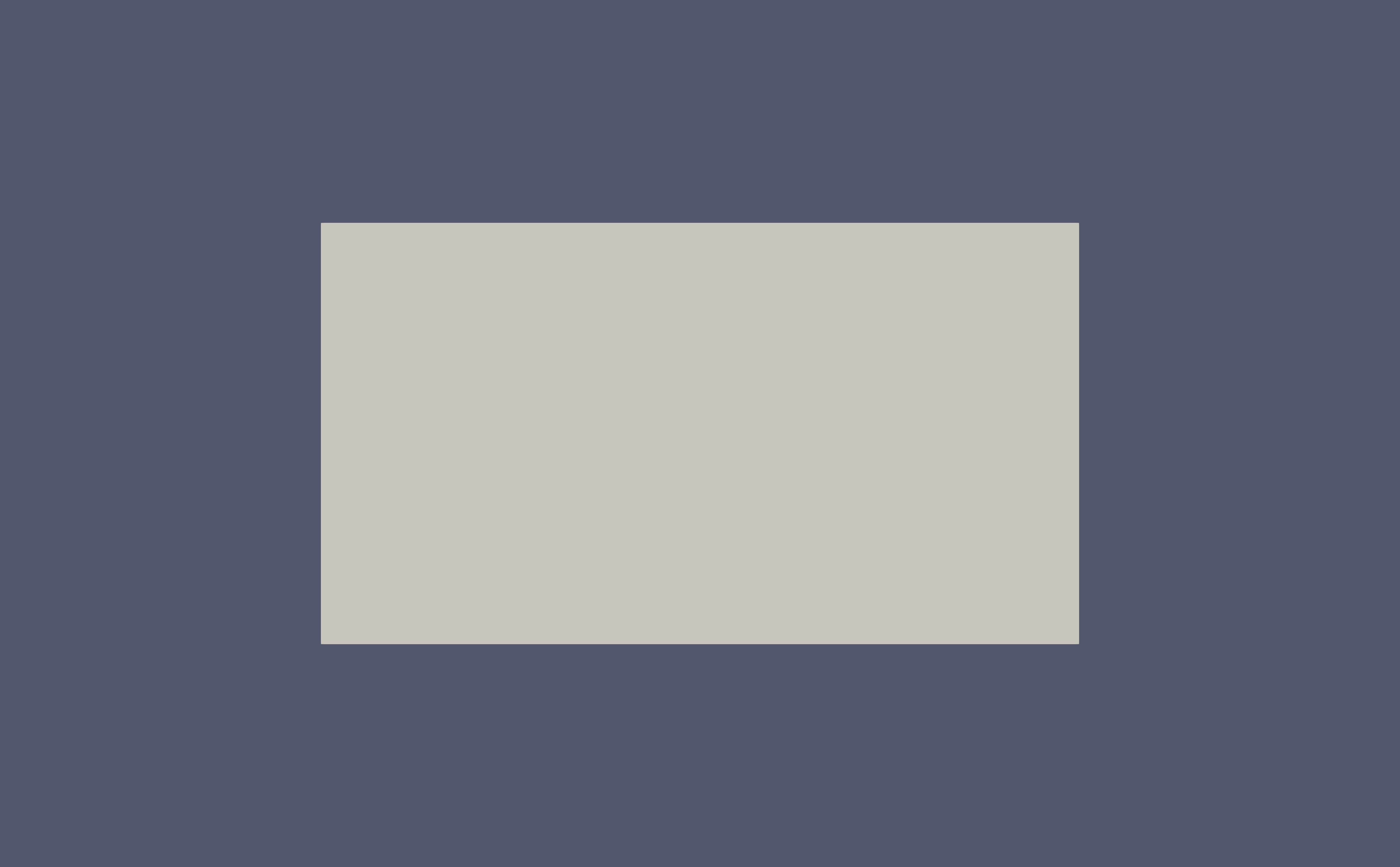}
\caption{Geometries for island (\emph{left}) and house (\emph{right}) represented by polygons.}
\label{figgeos}
\end{figure}

\begin{figure}[H]
\centering
\includegraphics[width=0.373\linewidth,keepaspectratio]{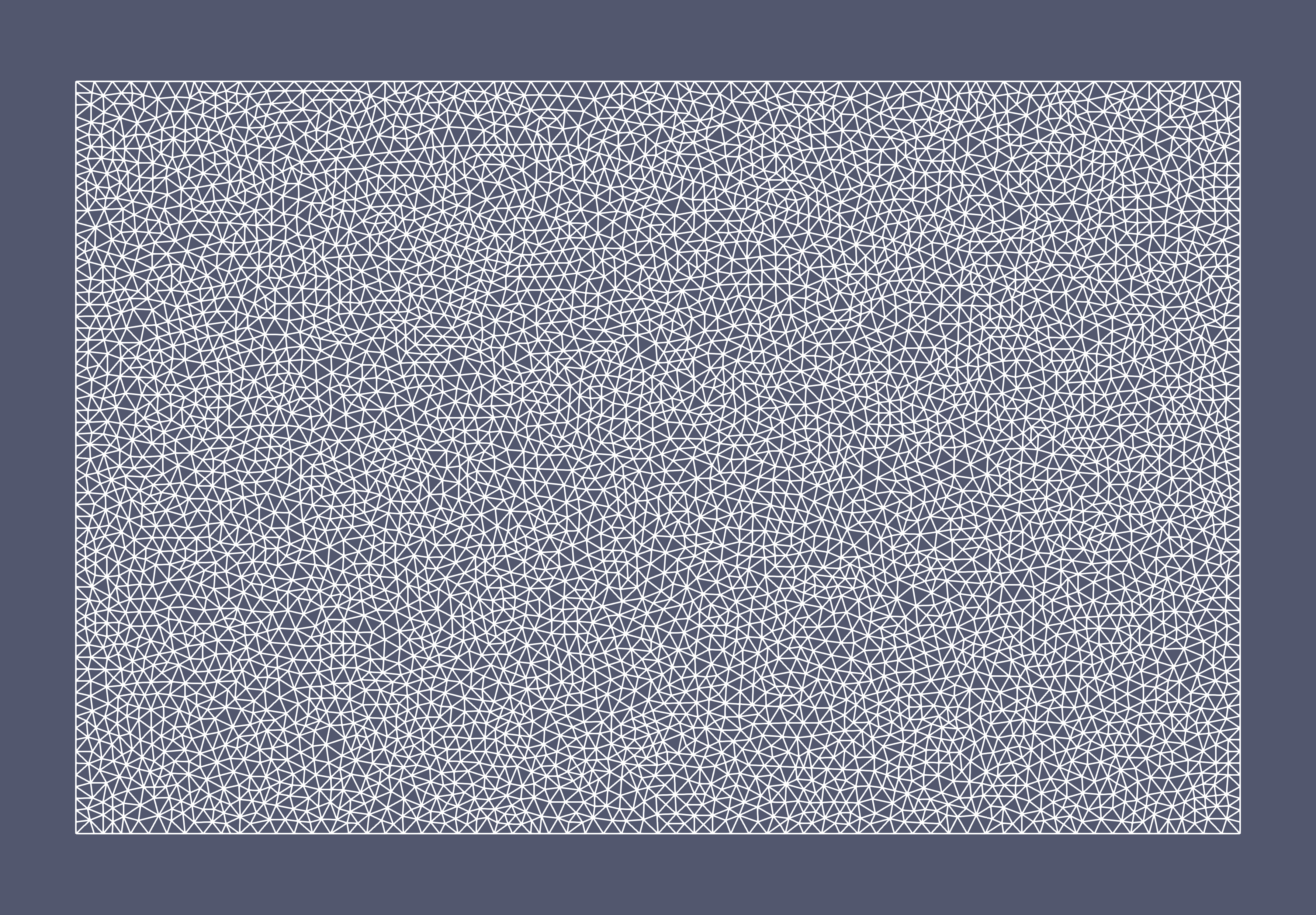}
\includegraphics[width=0.3045\linewidth,keepaspectratio]{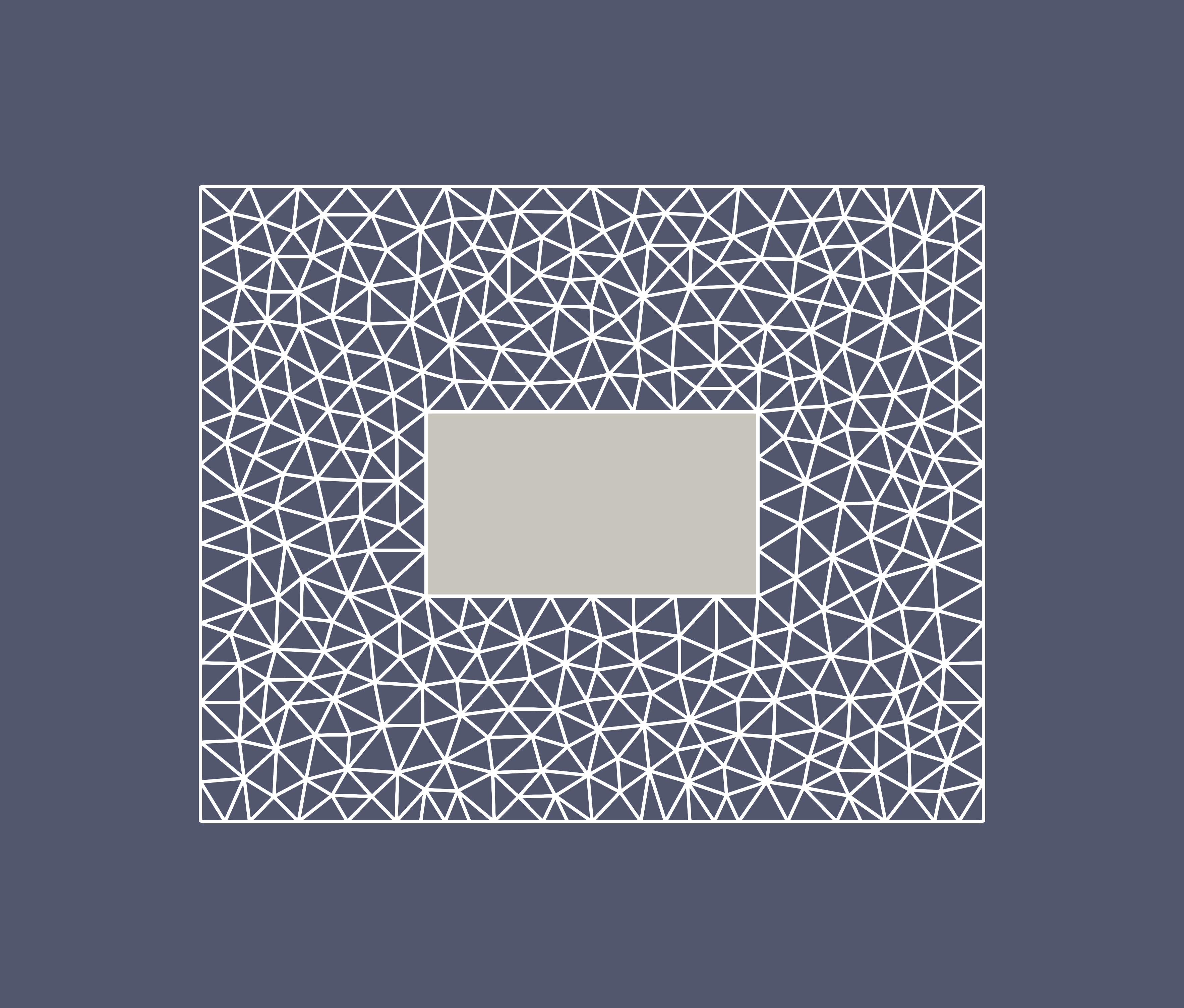}
\includegraphics[width=0.3045\linewidth,keepaspectratio]{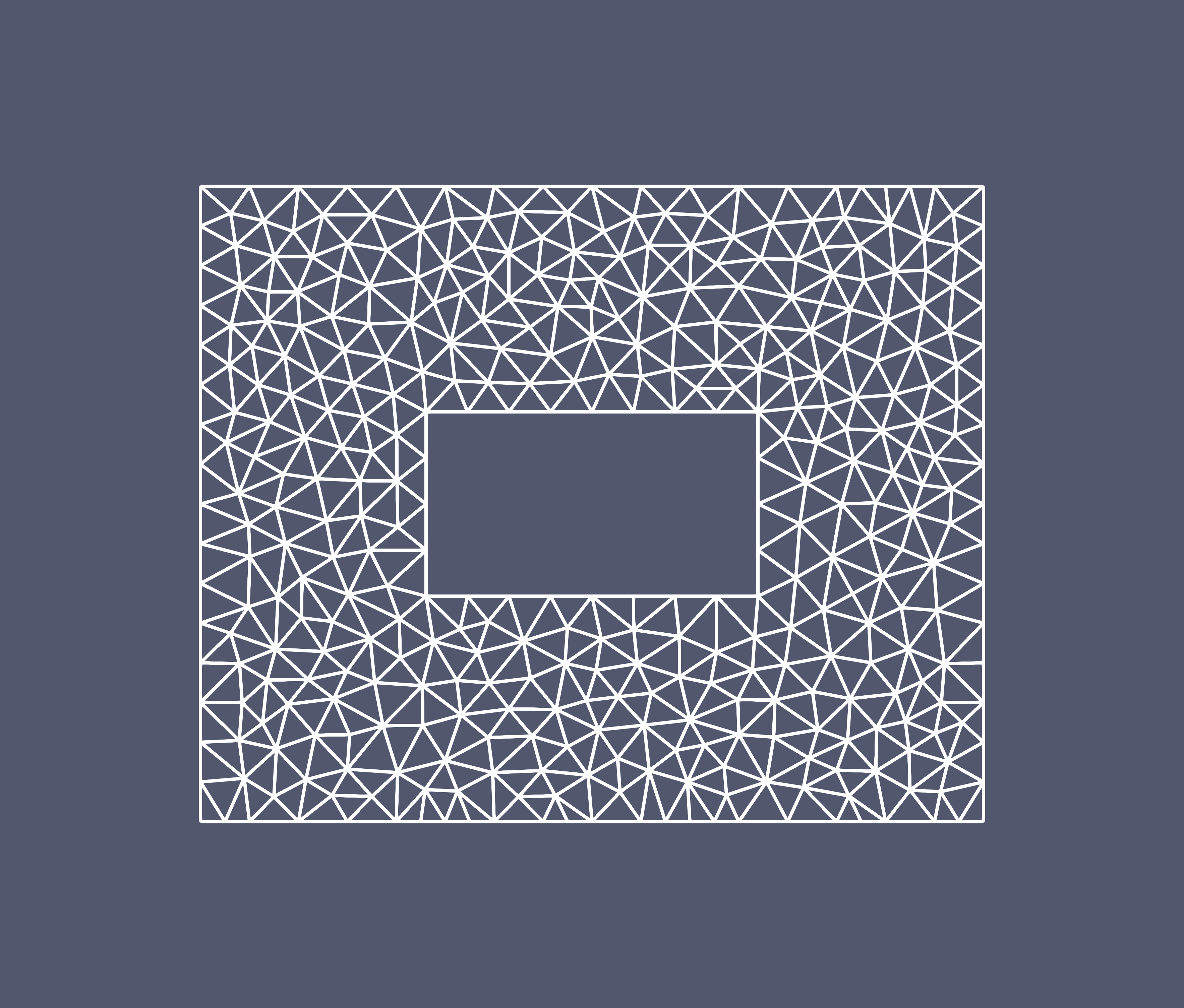}
\caption{Meshes generated from geometries. \emph{Left}: Air mesh with embedded boundary of island geometry. \emph{Middle}: House mesh around house geometry. \emph{Right}: House mesh without house geometry. Note the house-shaped hole in the mesh.}
\label{figmeshes}
\end{figure}

\begin{figure}[H]
\centering
\includegraphics[width=0.373\linewidth,keepaspectratio]{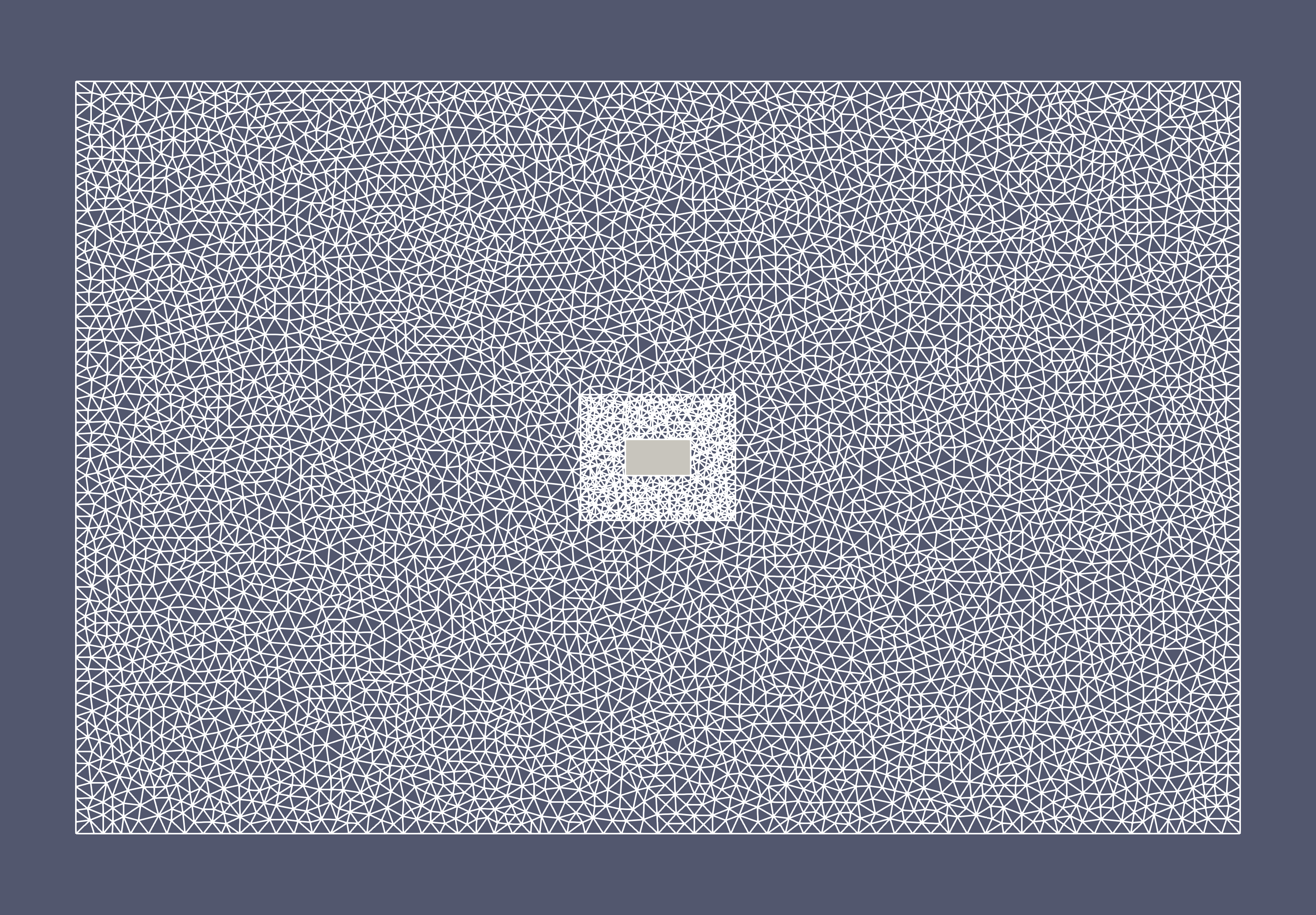}
\includegraphics[width=0.3045\linewidth,keepaspectratio]{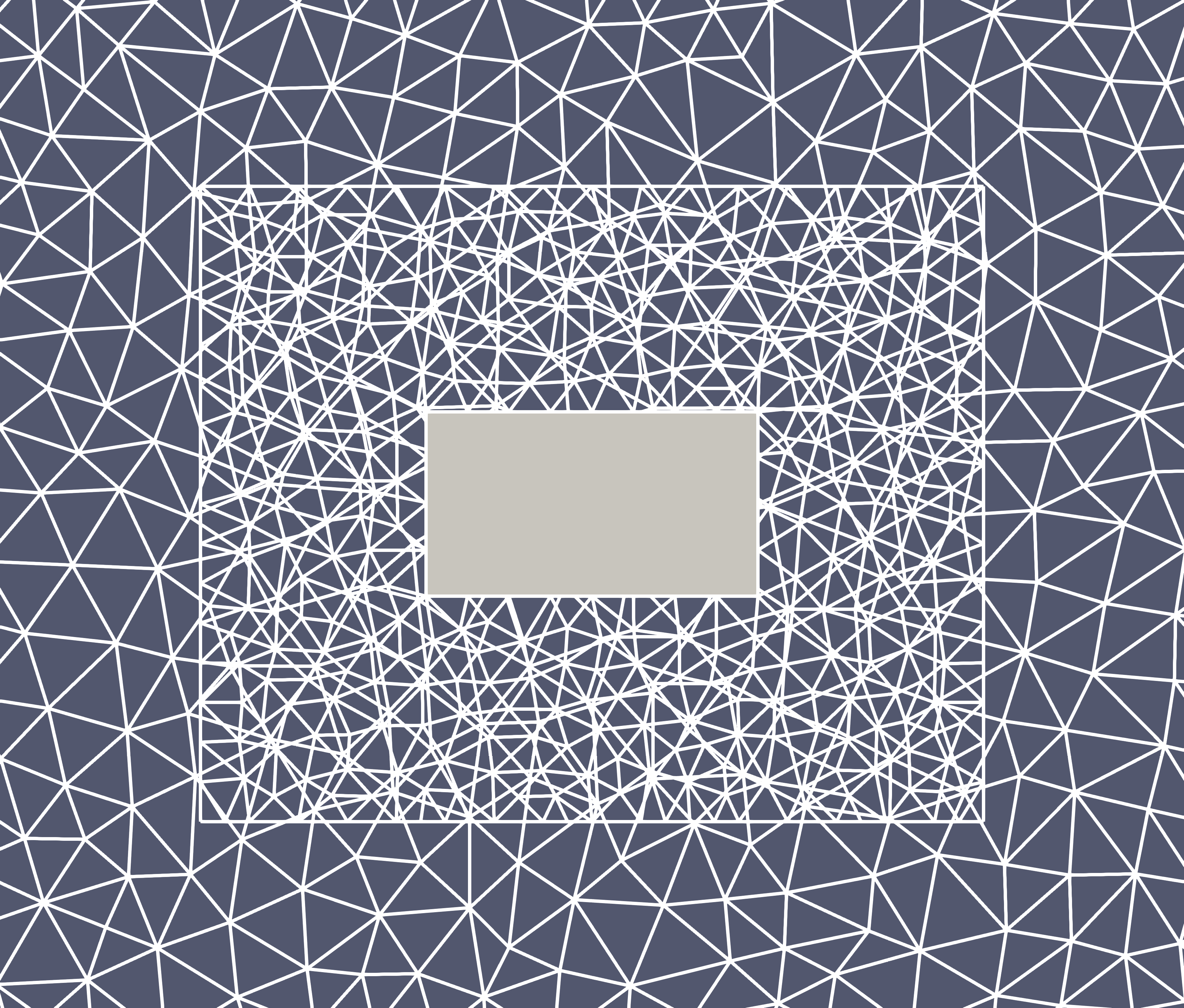}
\includegraphics[width=0.3045\linewidth,keepaspectratio]{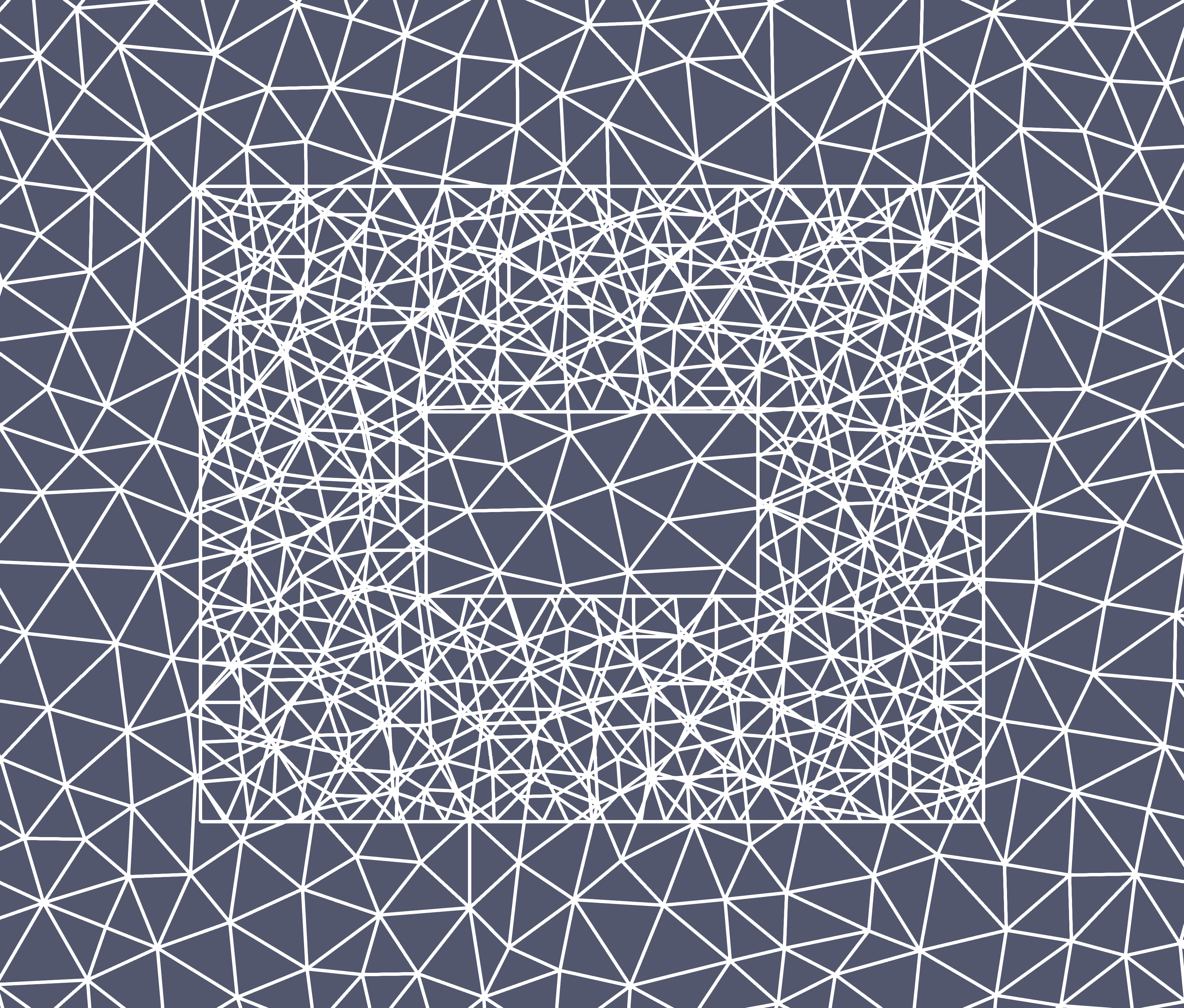}
\caption{Placement of overlapping house mesh in background air mesh with and without house geometry. Note the presence of background mesh cells in the hole of the overlapping mesh.}
\label{figmeshoverlap}
\end{figure}

\begin{figure}[H]
\centering
\includegraphics[width=0.49\linewidth,keepaspectratio]{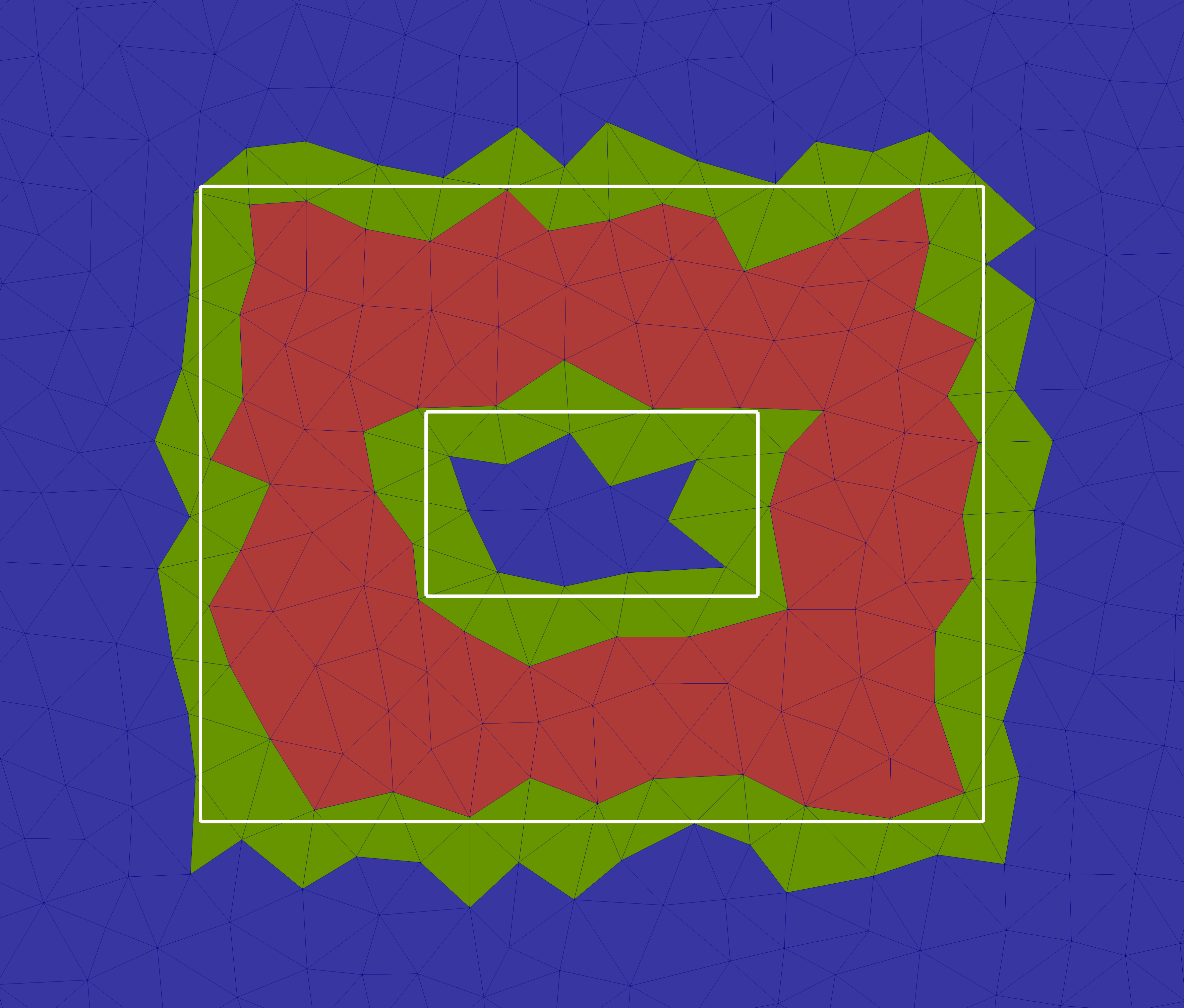}
\includegraphics[width=0.49\linewidth,keepaspectratio]{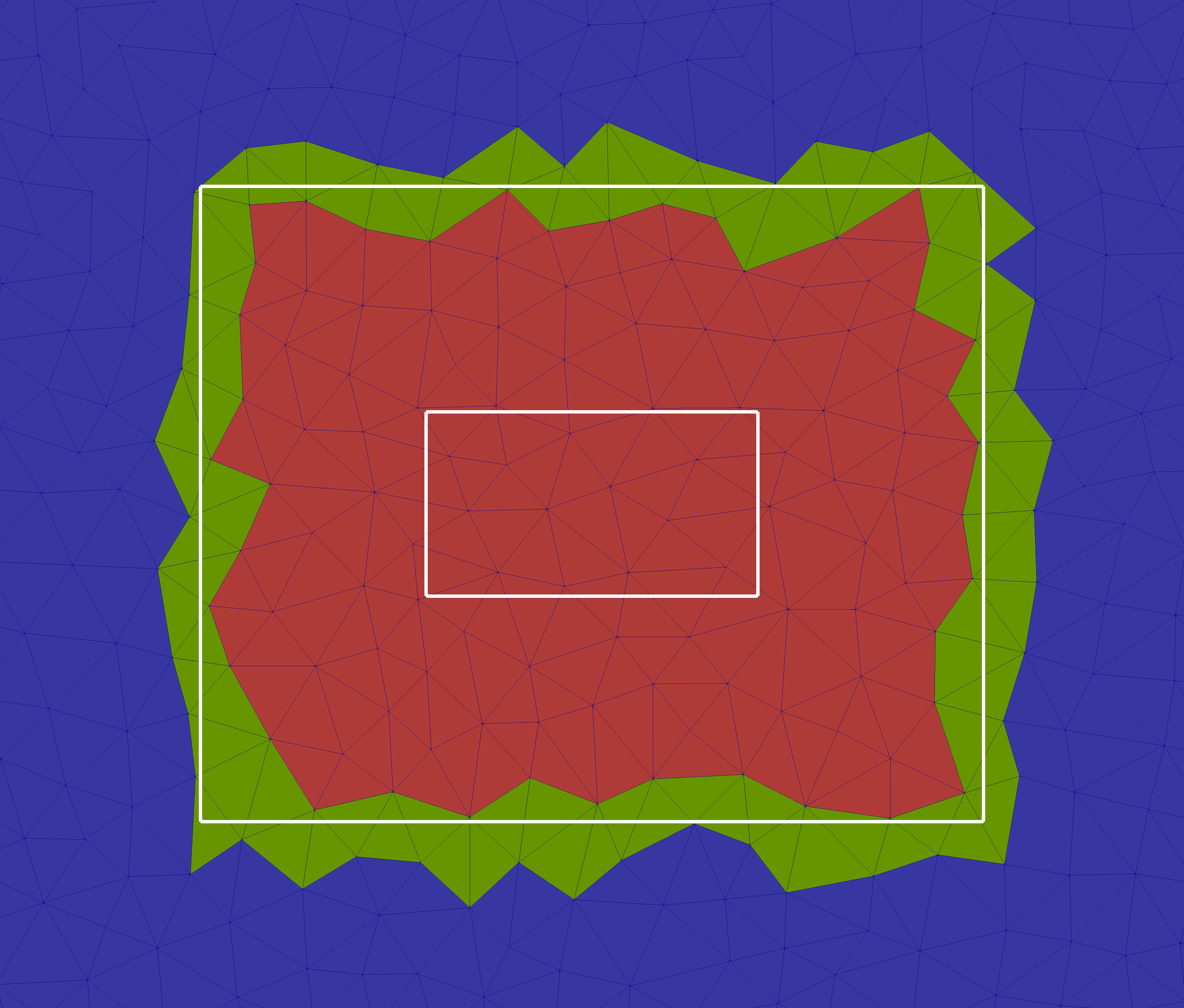}
\caption{Classification of background mesh cells before (\emph{left}) and after (\emph{right}) locating and marking cells intersecting the hole of the overlapping mesh outlined in white. Uncovered cells are blue, cut cells are green, and covered cells are red.}
\label{figautocoverdemo}
\end{figure}

\subsection{Computation of view}

It is well known that the view from a real estate can
affect the value of the property. However, the impact of view on real estate prices
has not been researched as much as other factors. Examples from literature, that
discuss the value of view are \cite{HUSSAIN2014316, Benson1998, Bourassa2004}.
However, these papers classify the view based on subjective human opinion.
In \cite{Benson1998} view is classified by different types: ocean, lake or mountain view,
where some of them are further classified into four subclasses from full ocean view to partial ocean
view. It is then estimated how large impact the different types of view have on the
price. Instead of a purely subjective classification, we in this paper present a novel and quantitative
measure, denoted $V$, for evaluating the view from a location such as the window of a building.
The measure gives a value between zero and one, zero being the worst possible view and
one the best possible view. We present a more elaborate measure for 3D settings
and a quite basic measure for 2D settings, starting with the former.

It is interesting to first consider the worst and the best cases. For this
application, the worst case would occur if the view is nothing
else than another house. 
Thus in this case the view should evaluate to
$V \approx 0$. The best case, $V = 1$, would be a view consisting entirely of
sea and/or sky. 
Examining again the worst case, the distance to a neighboring house
should influence the value of the view. In general, the negative
impact of objects on the view should decrease by the distance.  Our
proposed measure of view is expressed as an integral over the
integration domain $\omega = \omega_\phi \times \omega_\theta$ of size
$|\omega|$, where $\omega$ represents the viewed surface:
%
%
\begin{equation}
  V := \frac{1}{|\omega|} \iint \limits_\omega \sigma(\phi, \theta) \diff \phi \diff \theta \ .
  \label{eq:V}
\end{equation}
Here, the function $\sigma$ is the view density. It depends on the angles $\phi$ and $\theta$, and it
should take a value between zero and one. In the present study, we have used the following definition:
\begin{equation}
  \sigma(\phi, \theta) := \left\{
    \begin{array}{ll}
      1, & \text{if water or sky}, \\
      2 S\left( w(\phi,\theta) \frac{l(\phi,\theta)}{L} \right) -1, & \text{otherwise},
    \end{array} \right.
\end{equation}
where $S$ is the Sigmoid function, $S(t) = (1+\exp(-t))^{-1}$;
$w(\phi,\theta)$ is the specific weight for the object viewed at $(\phi,\theta)$, and is set to be $0.1$ for house, and $0.7$ for ground;
 $l(\phi,\theta)$ is the distance to the nearest object viewed at $(\phi,\theta)$; and $L$ is a calibration parameter. Based on the premise that $\sigma(\phi, \theta) = 0.9$
for a house viewed at $(\phi, \theta)$ and placed at the horizon
approximately 5 kilometers away, the parameter $L$ is found to be
$0.217~\mathrm{km}$. As the size of objects decreases almost
exponentially with respect to the distance from the camera, $S(t)$
has the largest impact on objects nearby. The reason for choosing $2S(t) - 1$ instead of, e.g., the simpler $1 - \exp(-t)$, (both are $0$ for $t = 0$, and both go to $1$ when $t \rightarrow \infty$), is because of the slower growth of the former.

The measure defined in \eqref{eq:V} computes the view independent of cardinal
direction. However, in northern countries like Sweden, a southward view
is often weighted higher than a northward view because of the
sunlight. Thus, to obtain a view formula which can depend on the
cardinal direction, we let $\theta \in [0,2\pi]$ be the angle in the
horizontal plane. Let southward be $\theta = 0$, consequently
$\theta = \pi$ is northward. Assuming that one would weight
the southward view three times higher as the northward view,
the cardinal direction weight function could be expressed by
\begin{equation}
  D(\theta) = 1+\frac{1}{2}\sin(\theta-3\pi/2) \ .
\end{equation}
For a $360^{\circ}$ horizontal view valuation, we introduce
$D(\theta)$ in \eqref{eq:V}. We thus get the following
modified measure of view, which produces values in $[0, 1]$:
\begin{equation}
  V_{360} := \frac{1}{|\omega|} \int_0^{2\pi} D(\theta) \int_{\omega_\phi} \sigma(\phi, \theta) \diff \phi \diff \theta \ .
  \label{eq:V_360}
\end{equation}
%

\begin{figure}[H]
  \centering
  \includegraphics[width=0.7\linewidth,keepaspectratio]{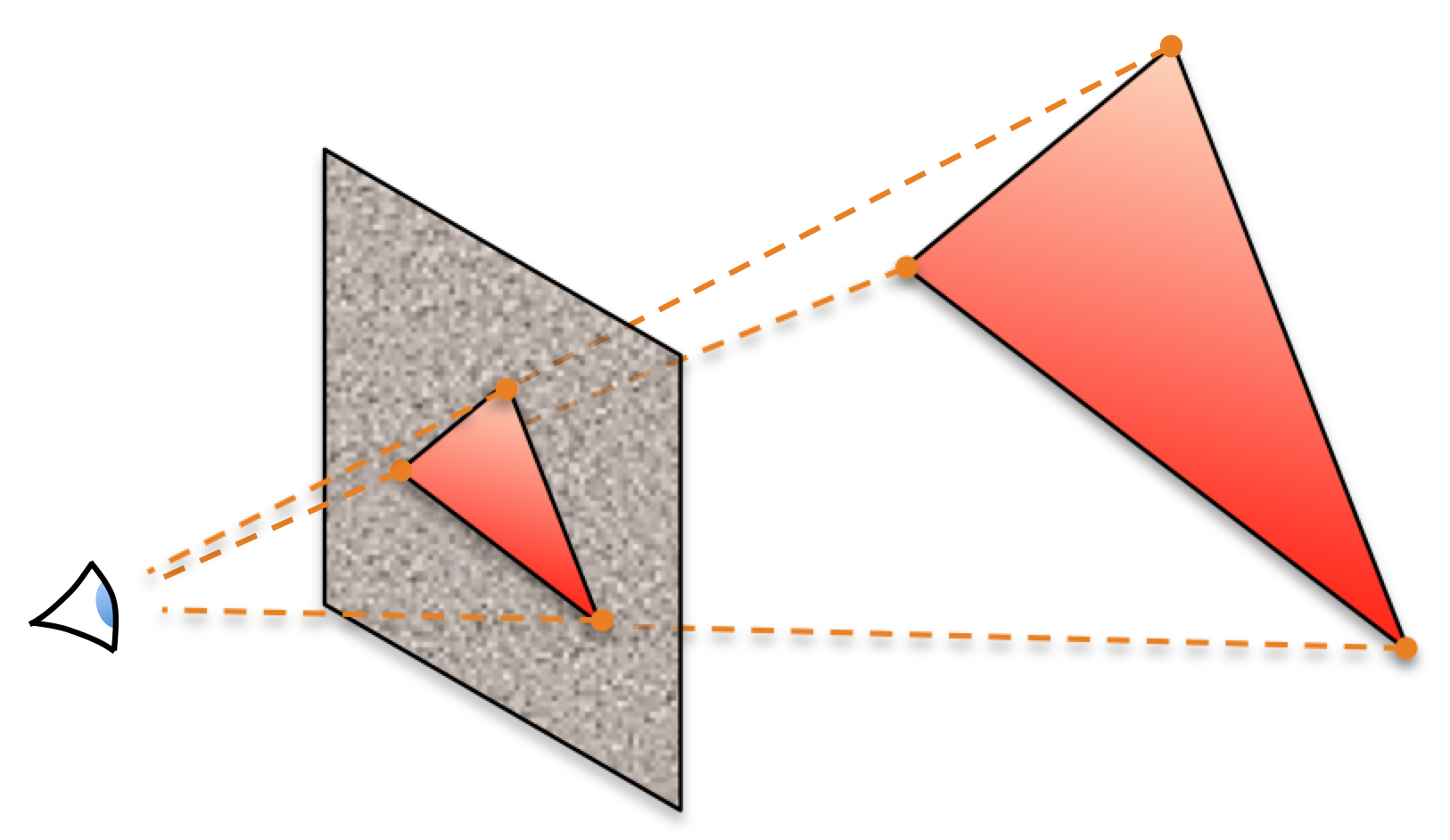}
  \caption{Projection of triangles from the 3D mesh onto the 2D image.}
  \label{fig:triangle-mapping}
\end{figure}

\begin{figure}[ht]
  \centering
  \includegraphics[width=0.46\linewidth,keepaspectratio]{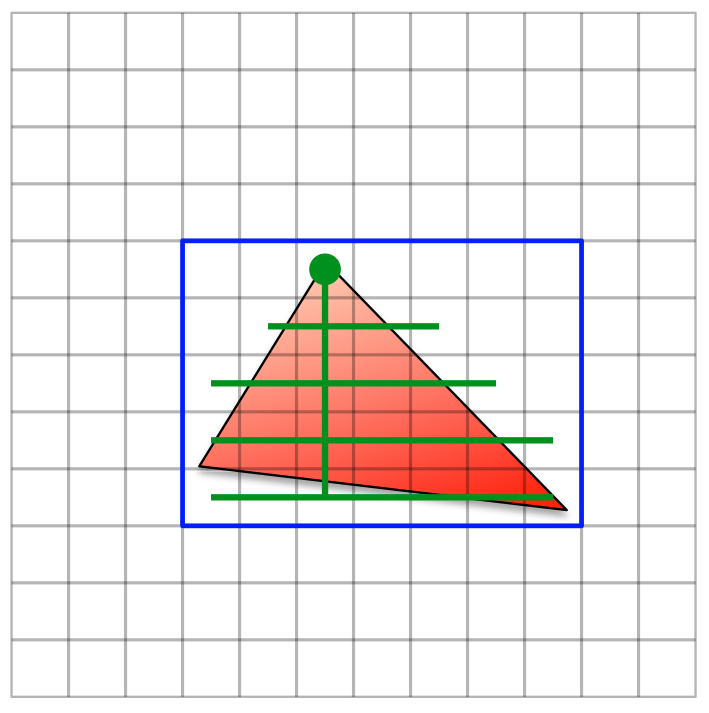}
  \caption{To check which pixels lie inside the triangle, we start
    from the top-point of the triangle (the green dot) going down one
    step at a time to check if the pixels to the left and right are
    contained in the triangle.}
  \label{fig:pixel-checking}
\end{figure}

Given the mesh(es) constructed for a 3D wind simulation, it is possible
to visualize the view from a given point. In practice we do this by
rendering an image from the 3D mesh with the use of the rasterization
rendering technique. This technique goes back to \cite{Catmull1974,
Pineda1988, Heckbert1989}. The rasterization algorithm projects the
triangles from the 3D mesh onto a 2D image. The projecting principle is
sketched in Figure \ref{fig:triangle-mapping}, where the vertices that are
mapped onto the 2D image plane are used to check which pixels the
triangle intersects.
\noindent The naive idea is to loop through all the pixels in the image
and check if they are inside the projected triangle or not. The
efficiency of this approach depends on the size of the triangles. To
account for small triangles, one may optimize the search by only
checking pixels that lie inside the bounding box of the
triangle. In Figure \ref{fig:pixel-checking}, the blue rectangle around
the triangle illustrates the bounding box for which the corner
coordinates are rounded off in order to include whole pixels.
This can be optimized further by not checking all the pixels inside
the bounding box, but starting in the pixel which contains the
top-point of the triangle. Then go stepwise down and check the pixels
to the left and right of the reference point. If we already visited
one or more pixels which are inside the projected triangle and then
comes to a pixel which is not in the triangle, the search stops in that
direction. The search algorithm is illustrated with the green lines in
Figure \ref{fig:pixel-checking}. There are several ways to optimize
this, see \cite{Pineda1988} for further reading. Looping over all the triangles
 to do the projections one by one can cause two or more projected
triangles to overlap. To decide which one that should be shown in the image,
we have to look at the distances to them. The distances to the triangles
that are already shown in the image are stored in a two dimensional
array with the same dimension as the image. Thus only the triangle
with the shortest distance may be shown in the image.

The rasterization algorithm is implemented in C++ and by the use of
the SWIG interface compiler it is possible to access the rasterization
algorithm from a Python script. The rasterization
algorithm needs as input a list of FEniCS (.xml) meshes, where the
first mesh in the list should be the background air mesh, and the rest
overlapping house meshes. With the scene set, the algorithm needs to know the size
of the image, both the size in pixels and the real size measured in
the same units as the meshes. Also the position and direction of the camera, and the
distance between the camera and the image are needed. With these
inputs, the algorithm generates an image and a matrix $\mathbf{S}$
of the same size as the image that contains a value for $\sigma$
for each pixel. The view $V$ may thus be computed from $\mathbf{S}$.
If we want to compute the $360^\circ$ view from
\eqref{eq:V_360}, the domain $\omega$ is assumed to be either a
cylinder or a sphere. As the rasterization algorithm generates
flattened images, \eqref{eq:V_360} cannot be used directly. However,
\eqref{eq:V_360} can be approximated by
\begin{equation} \label{eq:V360_estimate}
  V_{360} \approx \sum_{n=0}^{N-1} \frac{D\left( 2\pi n/N \right) }{N} V_n \ ,
\end{equation}
where $N\geq3$ is the number of images, and $V_n$ is the view valuation
for image number $n$, i.e., $V_n$ is evaluated
in the direction with angle $ 2\pi n/N$ (recall $0$ being south), and an
image width of $2d\tan(\pi/N)$, where $d$ is the distance between the
camera and the image. It is important that no images are overlapping
and that they, when joined together and projected onto the horizontal
plane, form the boundary of a convex regular polygon with $N$ edges.

We now present a simple view measure for 2D settings. Start by considering the projection of a settlement layout onto the horizontal plane. It does no longer make sense to consider the influence of ground, since the houses are immersed in it. Instead we let the view be either clear, taking $\sigma = 1$, or blocked by another house, taking $\sigma = 0$. The analogous 2D measure of (\ref{eq:V}), for some view point, is defined by
\begin{equation}
V^{2D} := \frac{1}{|\omega_\theta|}\int_{\omega_\theta} \sigma(\theta) \diff \theta.
\label{eq:V2d}
\end{equation}
\begin{wrapfigure}{r}{0.5\linewidth}
\centering
\includegraphics[width=0.5\textwidth,keepaspectratio]{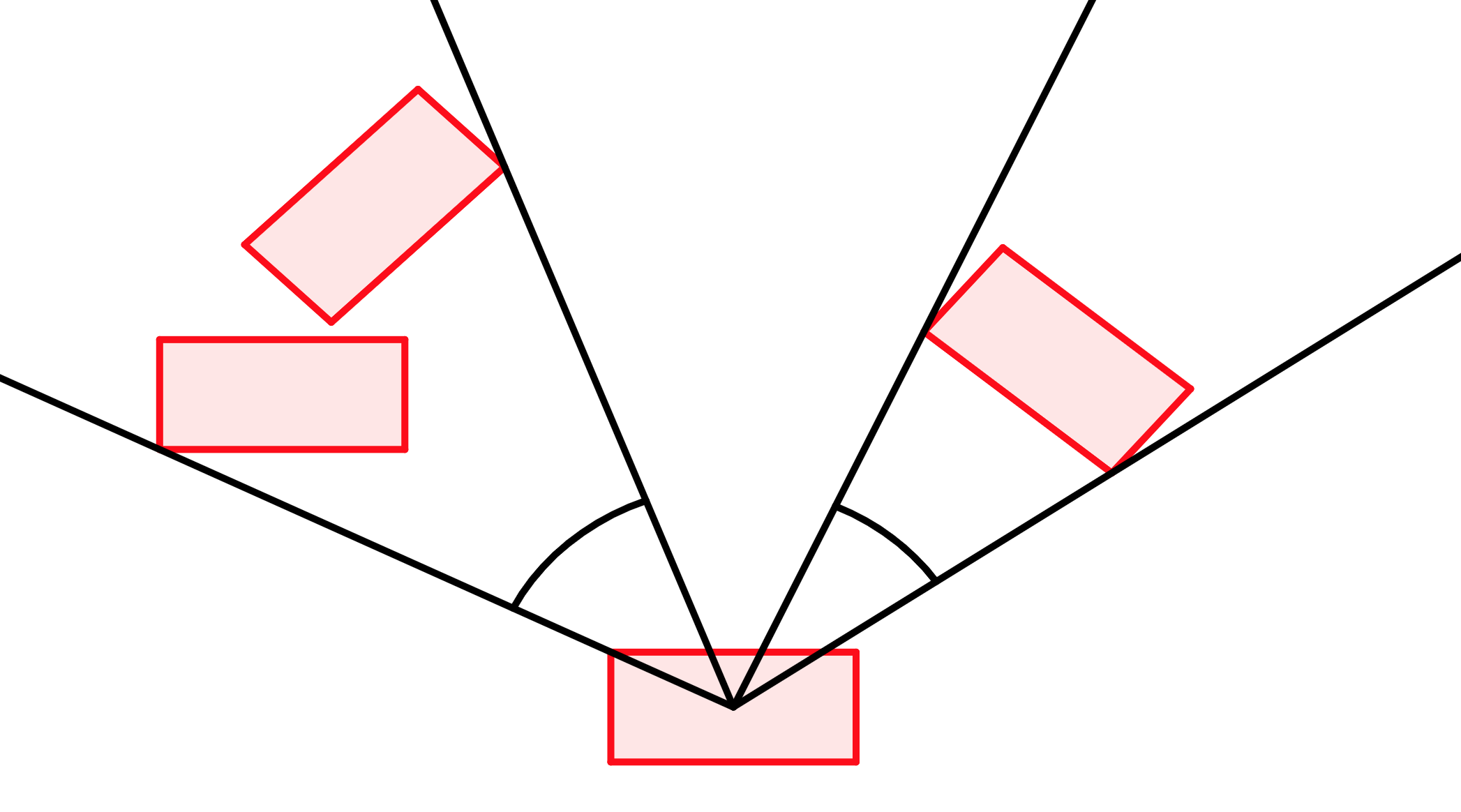}
\caption{View point taken as the midpoint of a house, resulting in two blocked circle sectors. Note that two or more houses may give rise to a single blocked circle sector.}
\label{figview2D}
\end{wrapfigure}
\noindent Due to the binary view density $\sigma$, (\ref{eq:V2d}) consists of clear and blocked circle sectors, see Figure \ref{figview2D}. The angle of a blocked circle sector is inversely proportional to the distance between the view point and the blocking house(s). Letting $\omega_\theta = [0, 2\pi)$, $i$ be a house index, and taking the view point to be the midpoint of house $i$, we may define the $360^\circ$ view from house $i$ by
\begin{equation}
V_{360, i}^{2D} := 1 - \frac{1}{2\pi} \sum_{j \in B_i} b_{ij},
\label{eq:Vi2d}
\end{equation}
\noindent where $B_i$ is the index set of blocked circle sectors for house $i$, and $b_{ij}$ is the angle in radians of blocked circle sector $j$ for house $i$. Note that (\ref{eq:Vi2d}) is equivalent to a simplified isovist.

\subsection{Multi-objective optimization}

The models for wind and view can be used to quantitatively evaluate a settlement layout. We may thus formulate a multi-objective optimization problem for finding the optimal placement of buildings in a landscape, such as houses on an island, with respect to the wind conditions around the buildings and the view from the buildings. Let $d = 2, 3$, be the spatial dimension of the settlement layout, and $H$ be an index set for the houses. The variables for the problem are the position and orientation of the houses:
\begin{equation}
\mbx = (\dots, x_{i1}, \dots, x_{id}, \theta_i ,\dots), \quad i \in H,
\label{eq:optvars}
\end{equation}
\noindent where $x_{i1}, \dots, x_{id}$ are the Cartesian coordinates for the midpoint of house $i$, and $\theta_i$ is the angle of rotation in the horizontal plane for house $i$ around its midpoint. To define an objective function, measures for both wind and view are needed. From construction, the output from the wind model is a function (velocity and pressure as a function of space) and not a number that says how good the wind is, as opposed to the view model. The natural mathematical way of measuring functions is to use norms. Thus, a measure of wind in a region $\Omega$ could be the $L^\infty(\Omega)$-norm of the velocity. In practice it can be more convenient to use an $L^p$-norm with a large $p$ instead of the $L^\infty$-norm, since this makes the objective function smoother. After trial and error testing in 2D settings, $p = 100$ was chosen for this study. Since wind can cause noise inside a house and damage the outside, we assume that it is desirable to have as weak wind as possible around all the houses. Hence we choose to minimize the wind around the house with the strongest surrounding wind. The collective measure of wind for a settlement layout is thus defined by
\begin{equation}
f_{W} := \max_{i \in H} \{ \| \mbu_h \|_{L^{100}(\Omega_i)} \}.
\label{eq:optfw}
\end{equation}
\noindent where $\mbu_h$ is the velocity field from (\ref{femstokes}), and $\Omega_i$ is the region of the overlapping house mesh for house $i$. The reason for choosing this region is simply that it is an already existing and easy accessible data structure for a surrounding region of a house, but one could of course work with other regions as well. For the view, we assume that it is desirable to have as good view as possible from all the houses. Hence, and for consistency with (\ref{eq:optfw}), we choose to minimize the complement of the $360^\circ$ view from the house with the worst view. The collective measure of view for a settlement layout is thus defined by
\begin{equation}
f_{V} := \max_{i \in H}  \{1 - V_{360, i} \},
\label{eq:optfv}
\end{equation}
\noindent where $V_{360, i}$ is the $360^\circ$ view for house $i$, defined by (\ref{eq:V_360}) for 3D and (\ref{eq:Vi2d}) for 2D. One may now use the two measures (\ref{eq:optfw}) and (\ref{eq:optfv}) to construct an objective function. In this study, we have taken it to be a convex combination of the two measures. This is more generally known
as linear scalarization with positive weights and minimizing such an objective function is
a sufficient but not necessary condition for Pareto optimality of the solution
according to, e.g., \cite{Marler2004}. The problem constraints are of two types: ``stay on island''-constraints, meaning that the houses must be placed on the island; and ``keep distance to neighbors''-constraints meaning that the houses may not be too close to each other. We denote the former by $s_i$, for $i \in H$, and the latter by $k_{ij}$, for $i,j \in H$ with $i < j$. Note that this is because the distance between houses $i$ and $j$ is the same as between houses $j$ and $i$, and that a house cannot keep a distance to itself. The multi-objective constrained problem for optimizing a settlement layout is thus
\begin{equation}
  \begin{split}
    \underset{\mbx}{\text{minimize}} \quad & \alpha f_W(\mbx) + (1-\alpha)f_V(\mbx) \\\
    \text{subject to} \quad & s_i(\mbx) \geq 0, \, i \in H, \\\
    & k_{ij}(\mbx) \geq 0, \, i, j \in H, i < j,
  \end{split}
  \label{eq:opt_prob_con}
\end{equation}
\noindent where the coefficient $\alpha \in [0, 1]$ is called the wind weight. Taking $\alpha = 0$ gives a pure view optimization problem, and $\alpha = 1$ pure wind.

The implementation is done with the Python package scipy.optimize, using the modified Powell algorithm. The choice of algorithm was based on a simple evaluation of the different methods available in scipy.optimize. The Powell method is also gradient-free, which is suitable considering the black box nature of the objective function in (\ref{eq:opt_prob_con}). However, constraints cannot be supplied to the Powell method in scipy.optimize, so they have to be added to the objective function, resulting in an unconstrained optimization problem. The reformulation of (\ref{eq:opt_prob_con}) as an unconstrained problem was taken to be
\begin{equation}
  \begin{split}
    \underset{\mbx}{\text{minimize}} \quad & \alpha f_W(\mbx) + (1-\alpha)f_V(\mbx) \\\
    + \, & p_s \sum_{i \in H} \min(0, s_i(\mbx))^2 + p_k \sum_{\substack{i,j \in H \\ i < j}} \min(0, k_{ij}(\mbx))^2,
  \end{split}
  \label{eq:opt_prob_uncon}
\end{equation}
\noindent where $p_s$ and $p_k$ are penalty parameters.

\section{Results}
\label{sec:results}

We present simulation results from the wind model in 2D, Section \ref{subsec:w2d}; the 3D view model, Section \ref{subsec:v3d}; the 2D view model, Section \ref{subsec:v2d}; and optimization of 2D settlement layouts, Section \ref{subsec:o2d}. Sections \ref{subsec:w2d} and \ref{subsec:v2d} are kept short, since Section \ref{subsec:o2d} naturally contains results from the wind model in 2D and the 2D view model.

\subsection{Wind 2D}
\label{subsec:w2d}

\begin{figure}[H]
\centering
\includegraphics[width=0.327\linewidth,keepaspectratio]{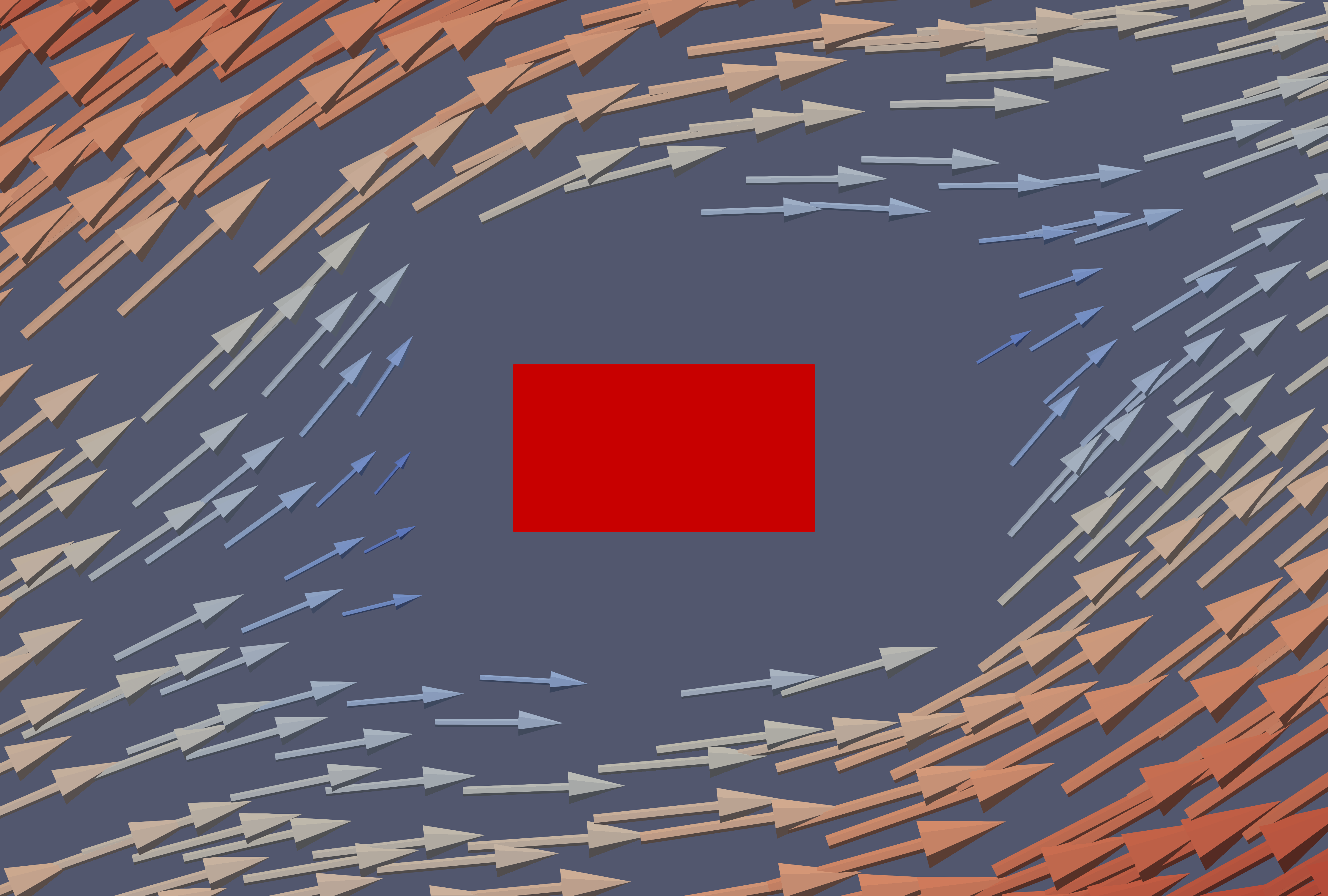}
\includegraphics[width=0.327\linewidth,keepaspectratio]{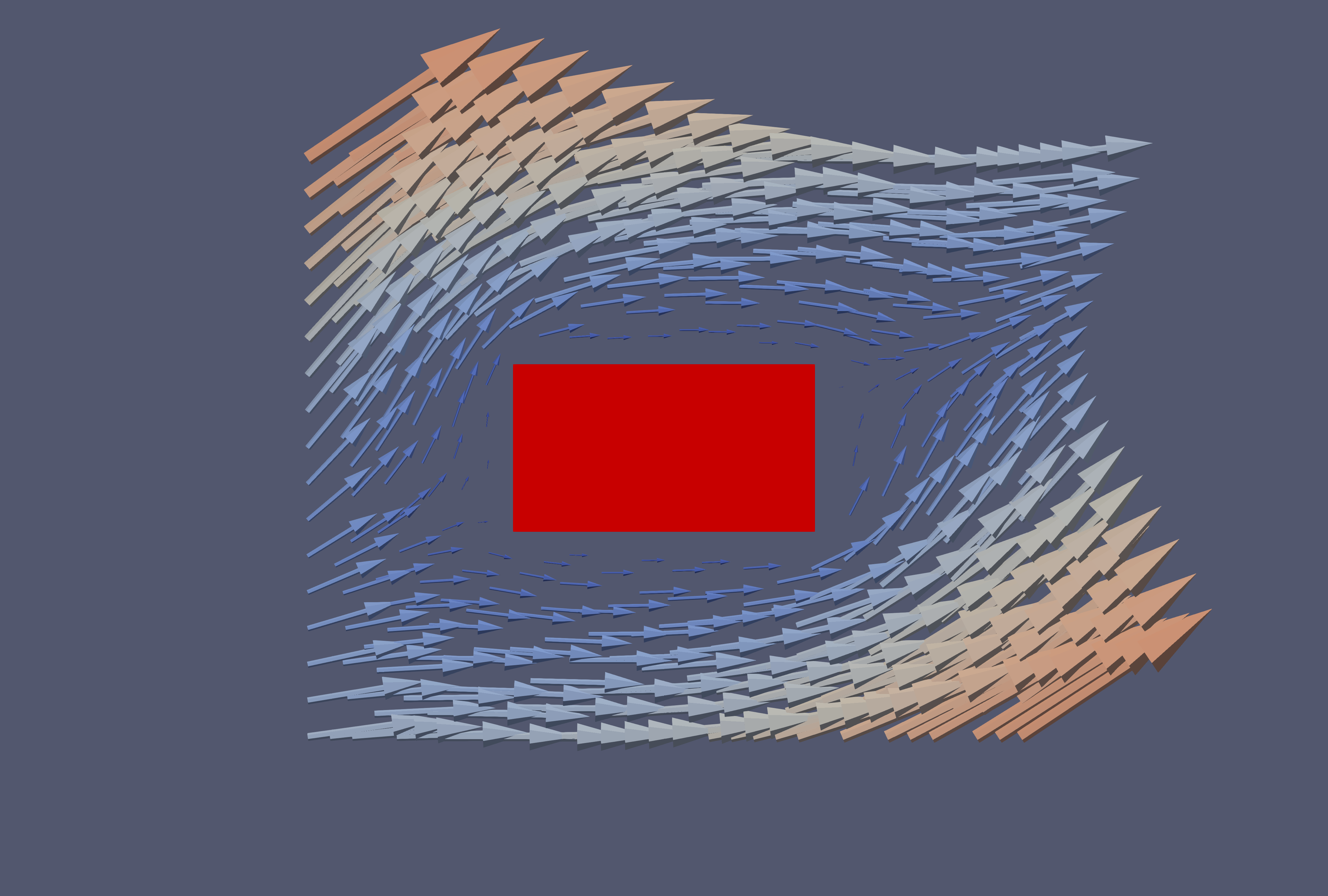}
\includegraphics[width=0.327\linewidth,keepaspectratio]{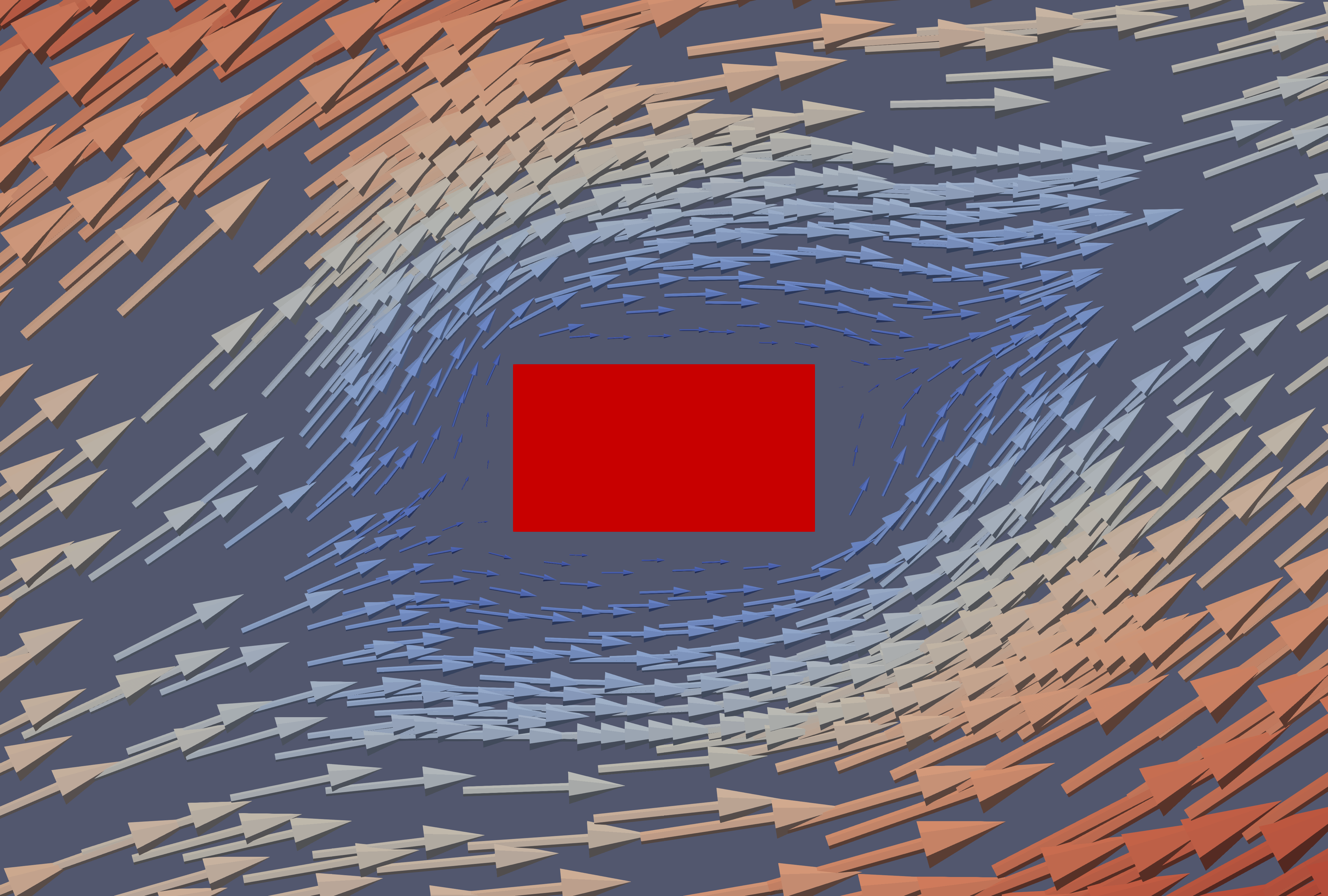}
\caption{\emph{Left}: Part of $\mbu_h$ from background air mesh. \emph{Middle}: Part of $\mbu_h$ from overlapping house mesh. \emph{Right}: Complete $\mbu_h$, i.e., from both meshes.}
\label{fig2Dflow}
\end{figure}

In Figure \ref{fig2Dflow}, the velocity, $\mbu_h$ from (\ref{femstokes}), for a 2D problem with one overlapping house mesh is shown. The velocity is visualized with glyphs, colored and scaled by magnitude. The house is shown in red

\subsection{View 3D}
\label{subsec:v3d}

\begin{figure}[H]
  \centering
  \includegraphics[width=\linewidth,keepaspectratio]{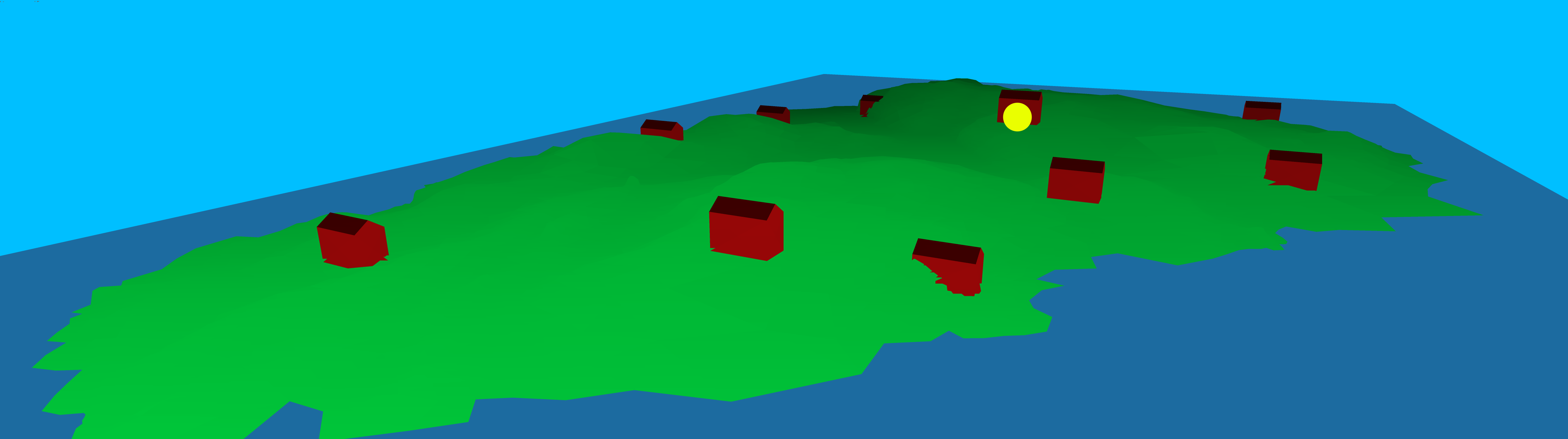}
  \caption{Overview of the island with ten houses. The yellow dot represents the position of the camera for the view computation.}
  \label{fig:camera-position}
\end{figure}

\begin{figure}[H]
\centering
\includegraphics[width=0.49\linewidth,keepaspectratio]{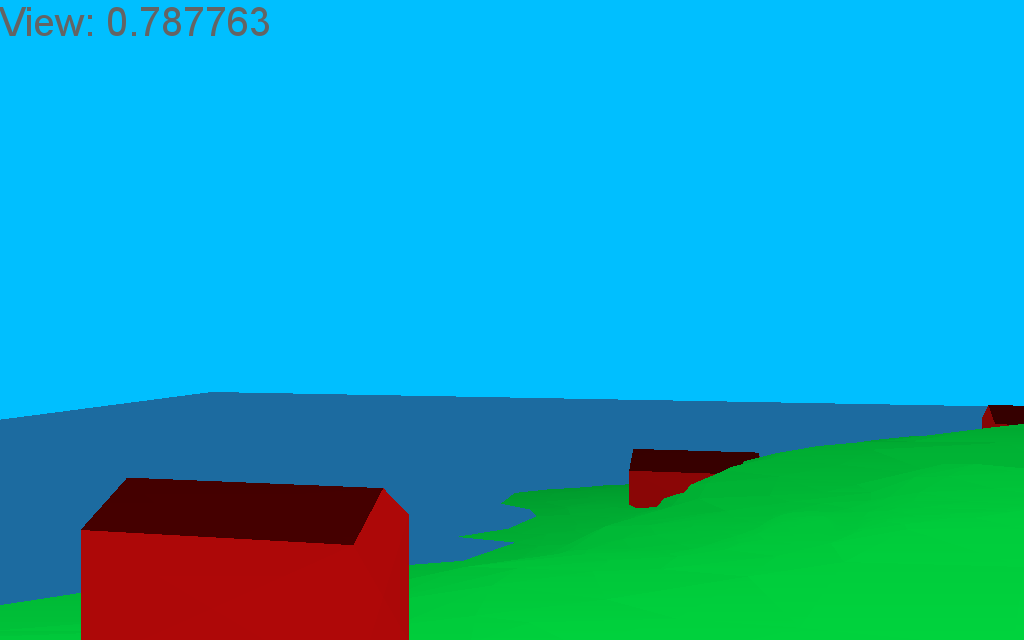}
\includegraphics[width=0.49\linewidth,keepaspectratio]{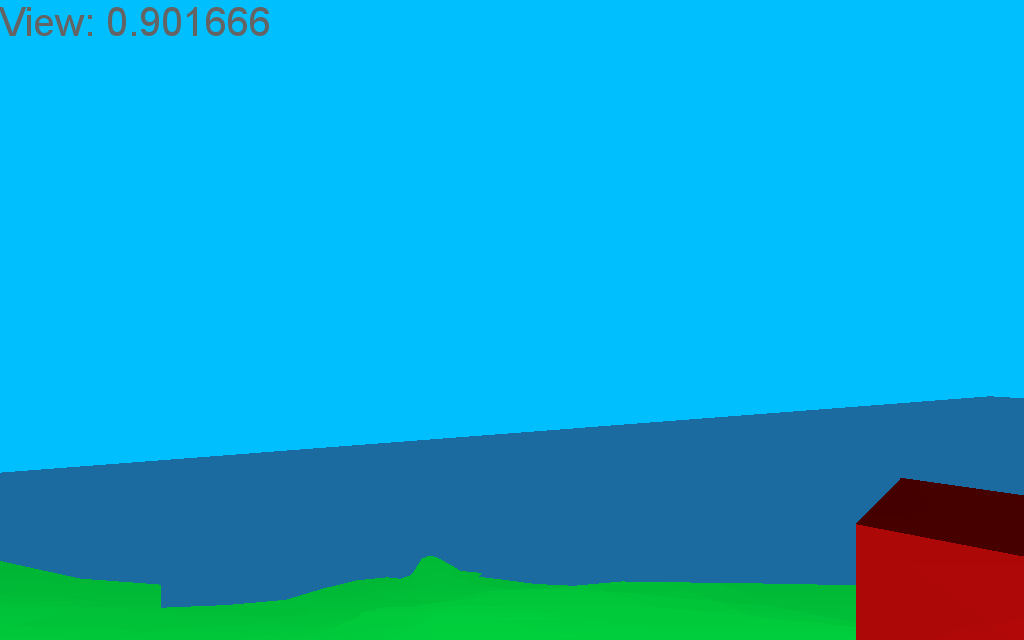}
\includegraphics[width=0.49\linewidth,keepaspectratio]{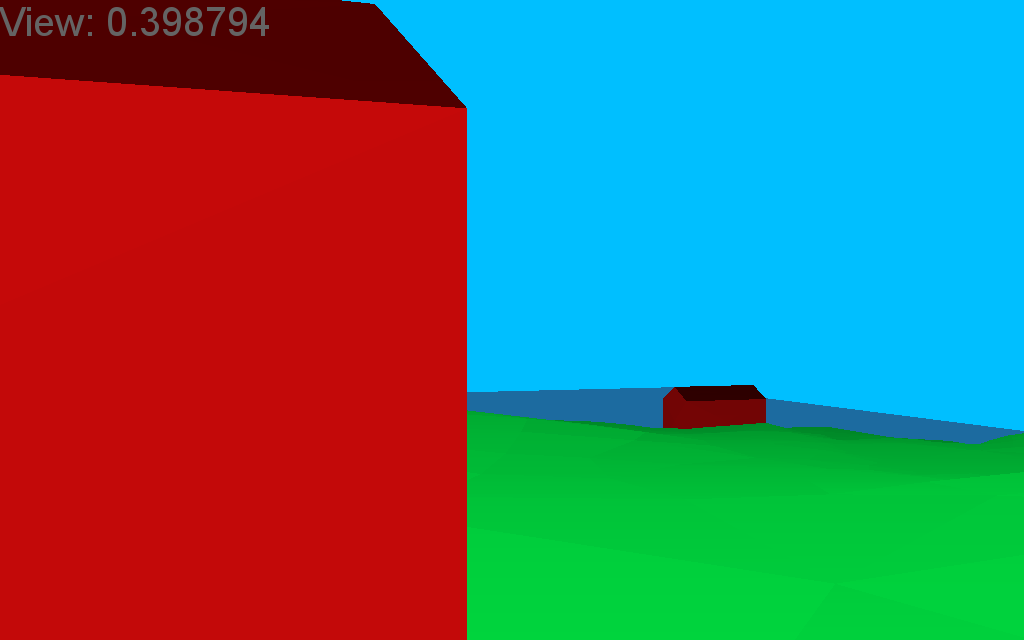}
\includegraphics[width=0.49\linewidth,keepaspectratio]{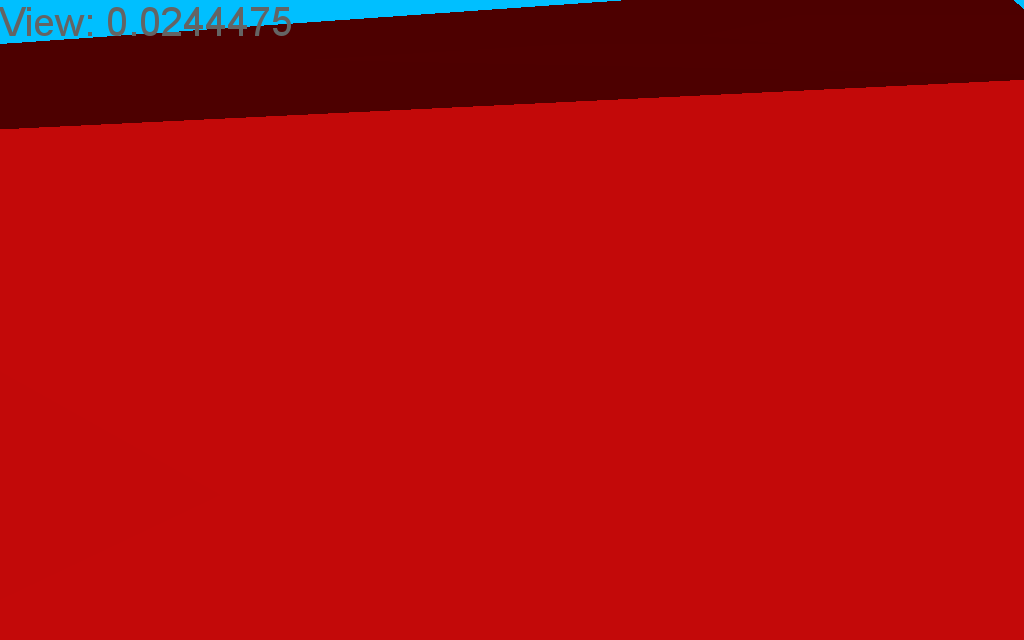}
\caption{\emph{Top left}: Fair view with $V = 0.79$. \emph{Top right}: Good view with $V=0.90$. \emph{Bottom left}: Poor view with $V = 0.40$. \emph{Bottom right}: Bad view with $V=0.02$}
\label{fig3Dview}
\end{figure}

To test the 3D view model, houses are arbitrarily placed on the island
as seen in Figure \ref{fig:camera-position}. The computed view from
the yellow dot in Figure \ref{fig:camera-position} for four different
directions is shown in Figure \ref{fig3Dview}.
\noindent The $360^{\circ}$ horizontal view measure $V_{360}$, defined by \eqref{eq:V360_estimate}, depends on the cardinal direction. For 32
computed rasterizations from the yellow dot in Figure \ref{fig:camera-position},
$V_{360} = 0.68$, if south is in the direction of the neighboring
house; and $V_{360} = 0.73$, if south is in the opposite direction.

\subsection{View 2D}
\label{subsec:v2d}

\begin{figure}[H]
\centering
\includegraphics[width=0.327\linewidth,keepaspectratio]{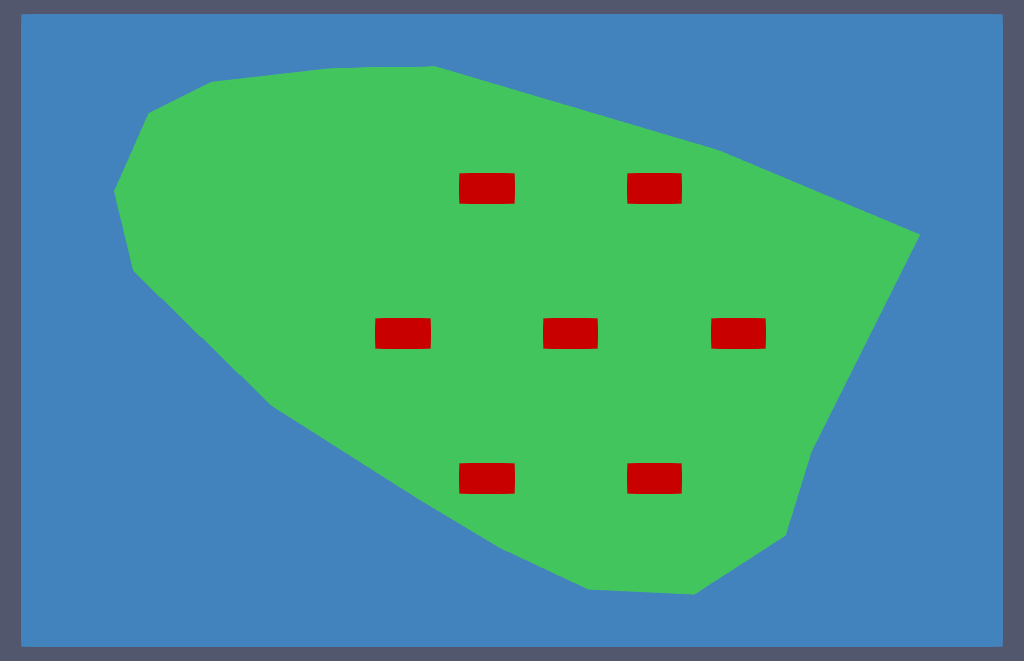}
\includegraphics[width=0.327\linewidth,keepaspectratio]{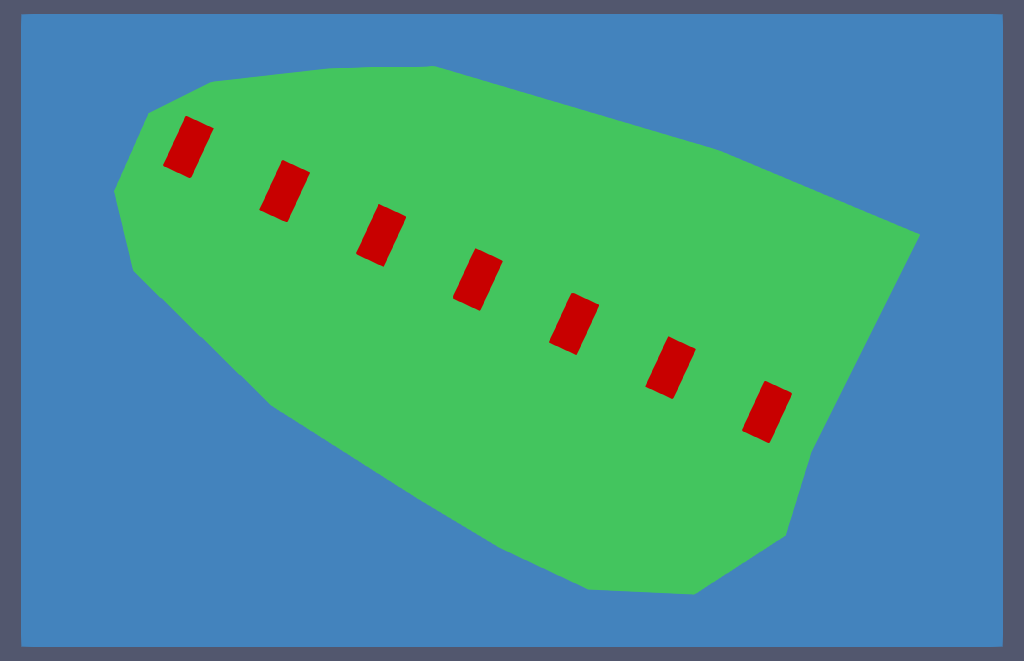}
\includegraphics[width=0.327\linewidth,keepaspectratio]{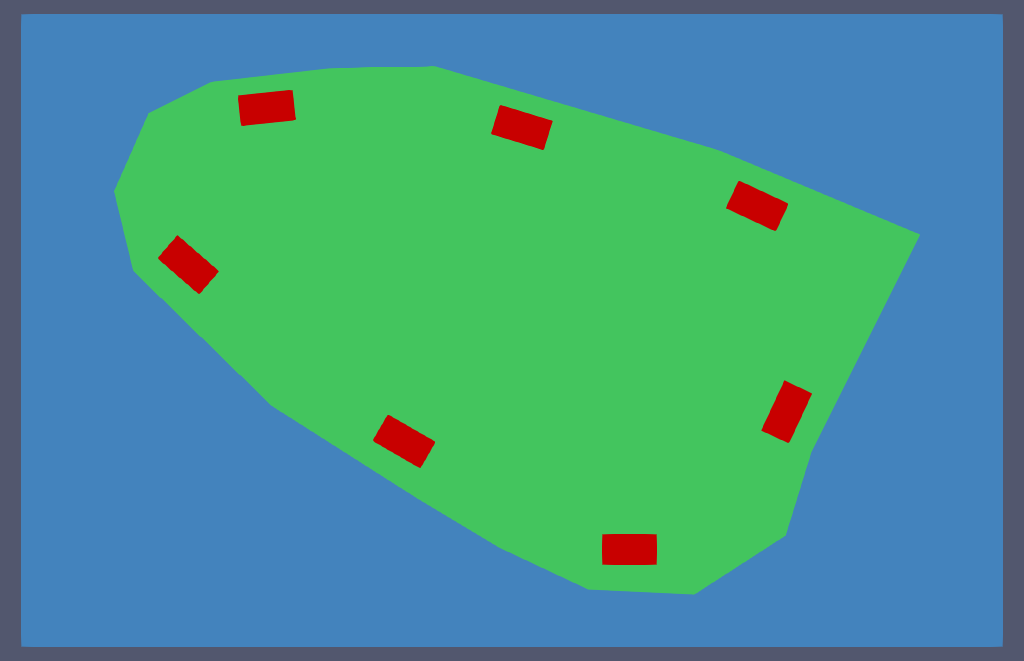}
\caption{\emph{Left}: Hexagonal layout with $f_V = \numprint{0.306355916815}$. \emph{Middle}: Principal line layout with $f_V = \numprint{0.188018240208}$. \emph{Right}: Coastal layout with $f_V = \numprint{0.167379980854}.$}
\label{fig2Dview}
\end{figure}

In Figure \ref{fig2Dview}, three different 2D settlement layouts with seven houses and different values for the collective measure of view $f_V$, defined by (\ref{eq:optfv}), are shown.

\subsection{Optimization 2D}
\label{subsec:o2d}

We present optimization results from solving problem (\ref{eq:opt_prob_uncon}) for 2D settlement layouts with seven houses, using the Powell method in scipy.optimize. The optimiziations are of three types: Pure wind ($\alpha = 1$), pure view ($\alpha = 0$), and mixed ($0 < \alpha < 1$). For all three types, three different initial settlement layouts are used: hexagonal, principal line, and coastal. They are shown in Figure \ref{fig2Dview}. The penalty parameters, chosen by trial and error, are $p_s =10^8$ and $p_k = 10^3$. For the wind simulations, seven identical copies of a house mesh are used for the houses. Together with the air mesh they constitute the multi-mesh hierarchy. See Table \ref{tab_mesh_data} for mesh data.
\begin{table}[H]
\begin{center}
\begin{tabular}{l | l l l}
    \multicolumn{4}{c}{Mesh data} \\ \hline
     & {Air mesh} & {House mesh} & {multi-mesh} \\ \hline
    {Number of vertices} &  5242 & 390 & 7972  \\
    {Number of edges} & 15453 & 1072 & 22957 \\
    {Number of cells} & 10212 & 682 & 14986 \\
\end{tabular}
\caption{Mesh data for background air mesh, house mesh, and multi-mesh hierarchy (background air mesh + 7 overlapping house meshes).}
\label{tab_mesh_data}
\end{center}
\end{table}
\noindent The system size is 69830, since Taylor-Hood elements are used, 3$\times$(number of vertices) + 2$\times$(number of edges). On a MacBook Pro with a 3.1 GHz Intel Core i7 processor, the average time for an evaluation of $f_W$ is 5.3s, and for $f_V$ 0.0086s (averaged over 100 simulations). The evaluation of $f_W$ thus takes approximately 600 times longer than that of $f_V$.

For pure wind optimization, three different inflow wind directions are used: horizontal (from left to right), vertical (from bottom to top), and diagonal (from lower left corner to upper right corner). The inflow wind speed is normalized to one. Numerical results from pure wind optimization in 2D are shown in Table \ref{tab_opt_purewind_specific}.
\begin{table}[H]
\begin{center}
\begin{tabular}{l l | n{1}{5}n{1}{5}n{1}{5}}
    \multicolumn{5}{c}{Pure wind optimization} \\ \hline
    {Wind $\setminus$ Layout} & & {Hexagonal} & {Principal line} & {Coastal} \\ \hline
    Horizontal & $f_W$ & 1.126550 {(\numprint{1.349040})} & 1.063120 {(\numprint{1.598102})} & 1.037030 {(\numprint{1.393428})} \\
                     & nfev & {3363} & {8307} & {12574} \\ \hline
    {Vertical} &  $f_W$ & 1.119143 {(\numprint{1.213518})} & 0.863266 {(\numprint{2.016172})} & 1.110130 {(\numprint{1.333130})} \\
                    & nfev & {1723} & {10119} & {6062} \\ \hline
    {Diagonal} & $f_W$ & 0.992282 {(\numprint{1.586989})} & 1.229826 {(\numprint{2.142663})} & 1.242318 {(\numprint{1.391669})} \\
                      & nfev & {13800} & {3185} & {2982} \\
\end{tabular}
\caption{Optimized values of the collective measure of wind, $f_W$, from pure wind optimizations. Starting values of $f_W$ are shown in parentheses. Number of evaluations of the objective function during optimization, nfev, are shown below.}
\label{tab_opt_purewind_specific}
\end{center}
\end{table}
%

\begin{figure}[H]
\centering
\includegraphics[width=0.327\linewidth,keepaspectratio]{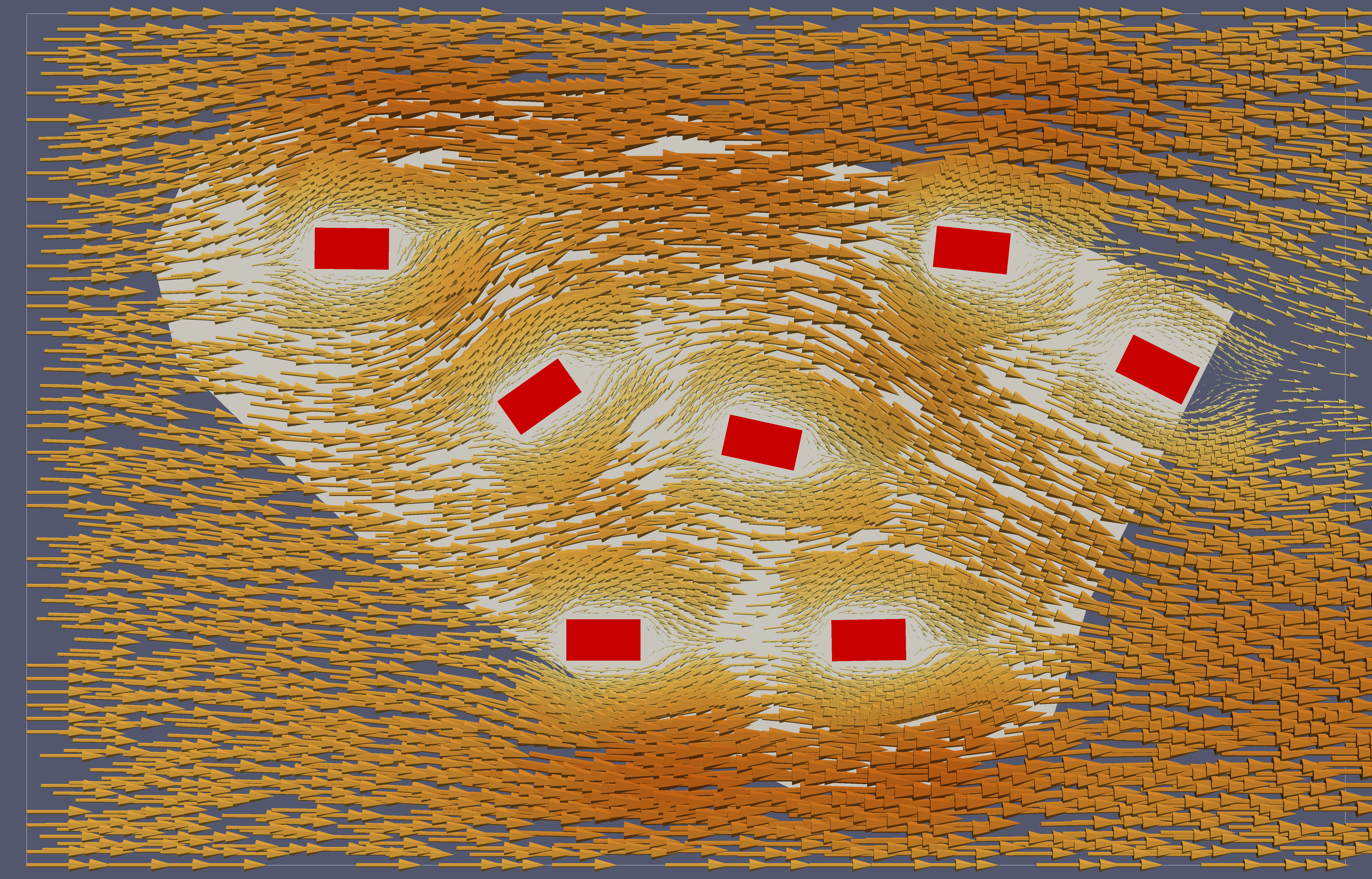}
\includegraphics[width=0.327\linewidth,keepaspectratio]{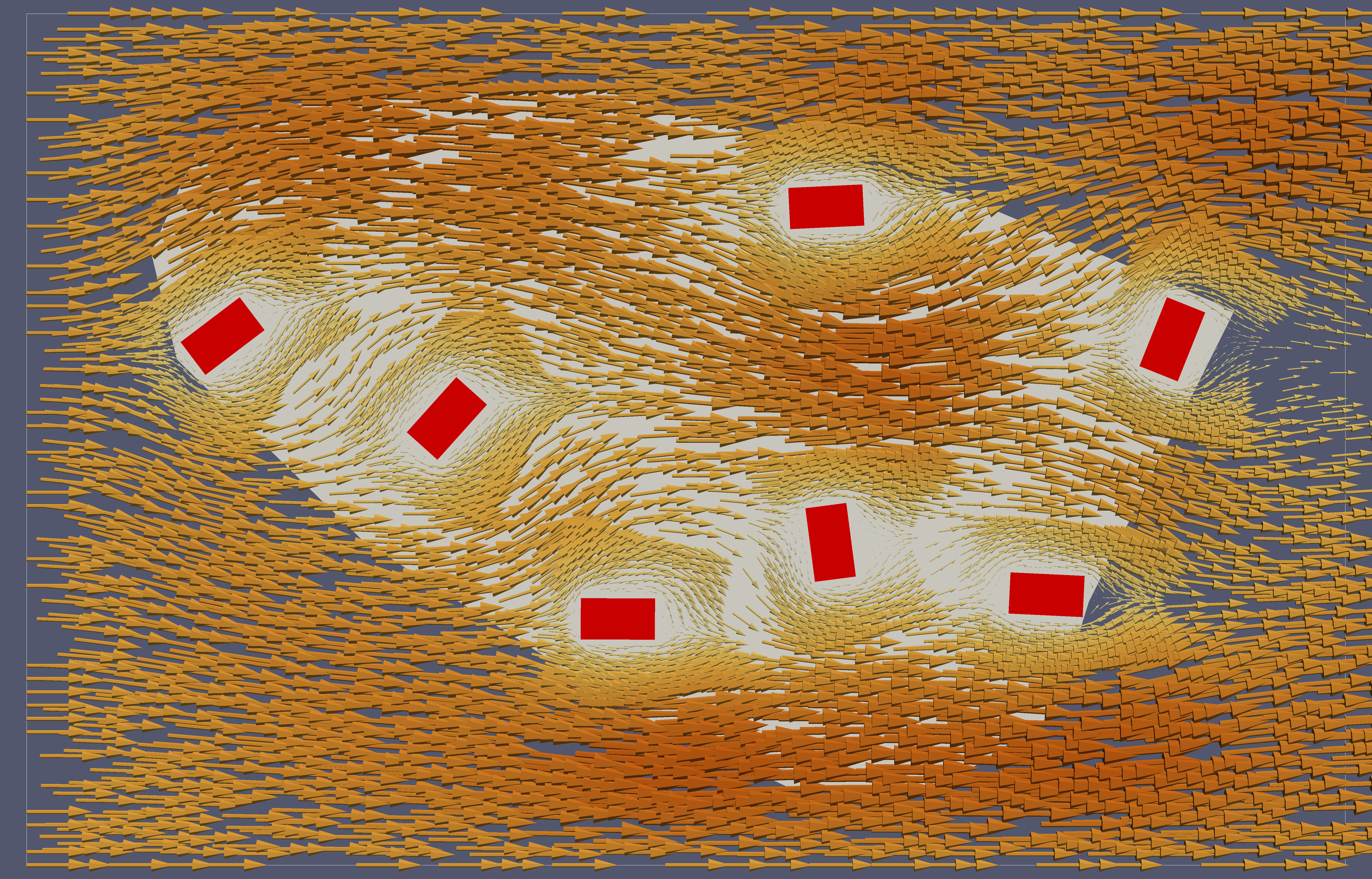}
\includegraphics[width=0.327\linewidth,keepaspectratio]{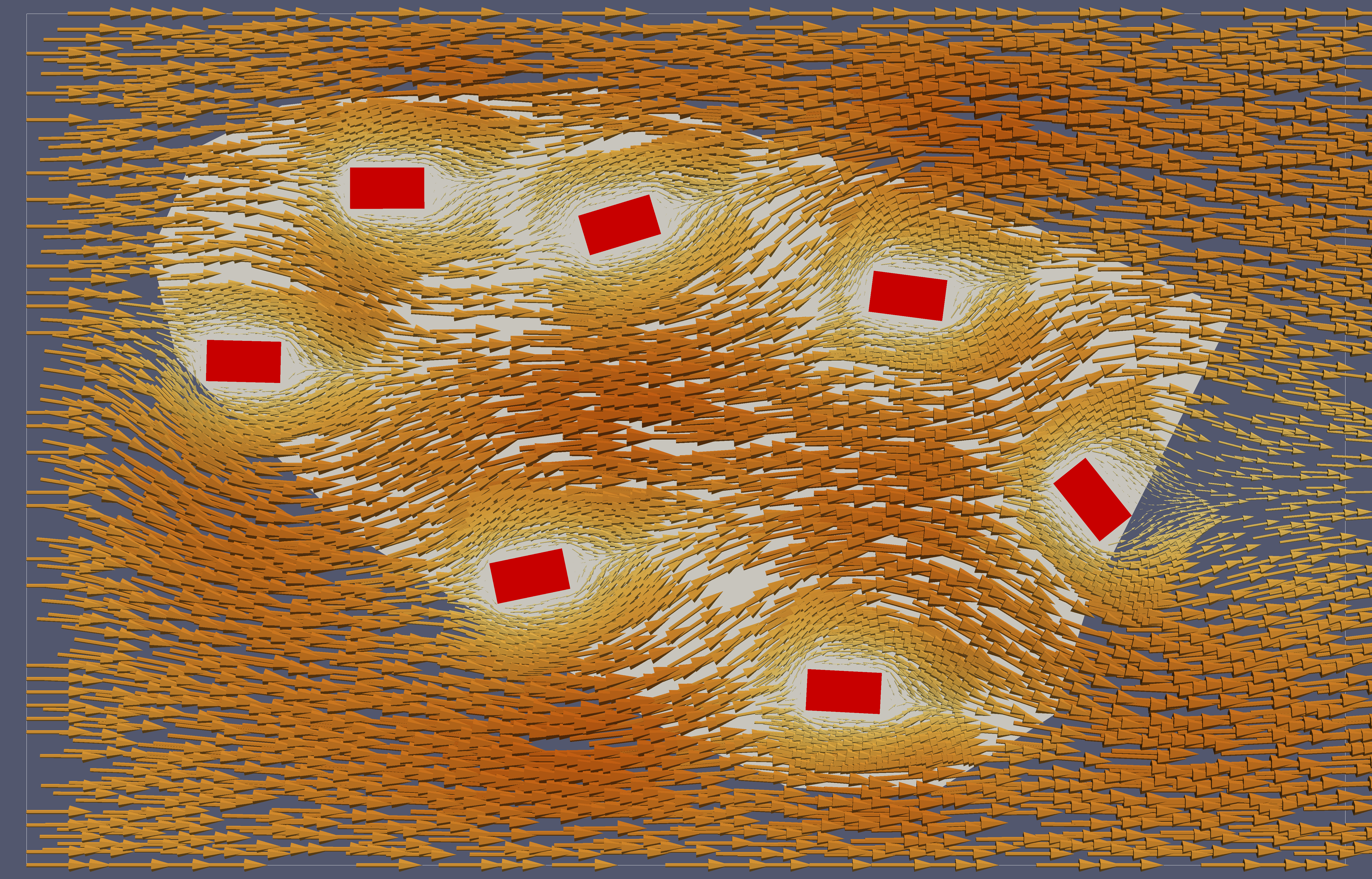}
\includegraphics[width=0.327\linewidth,keepaspectratio]{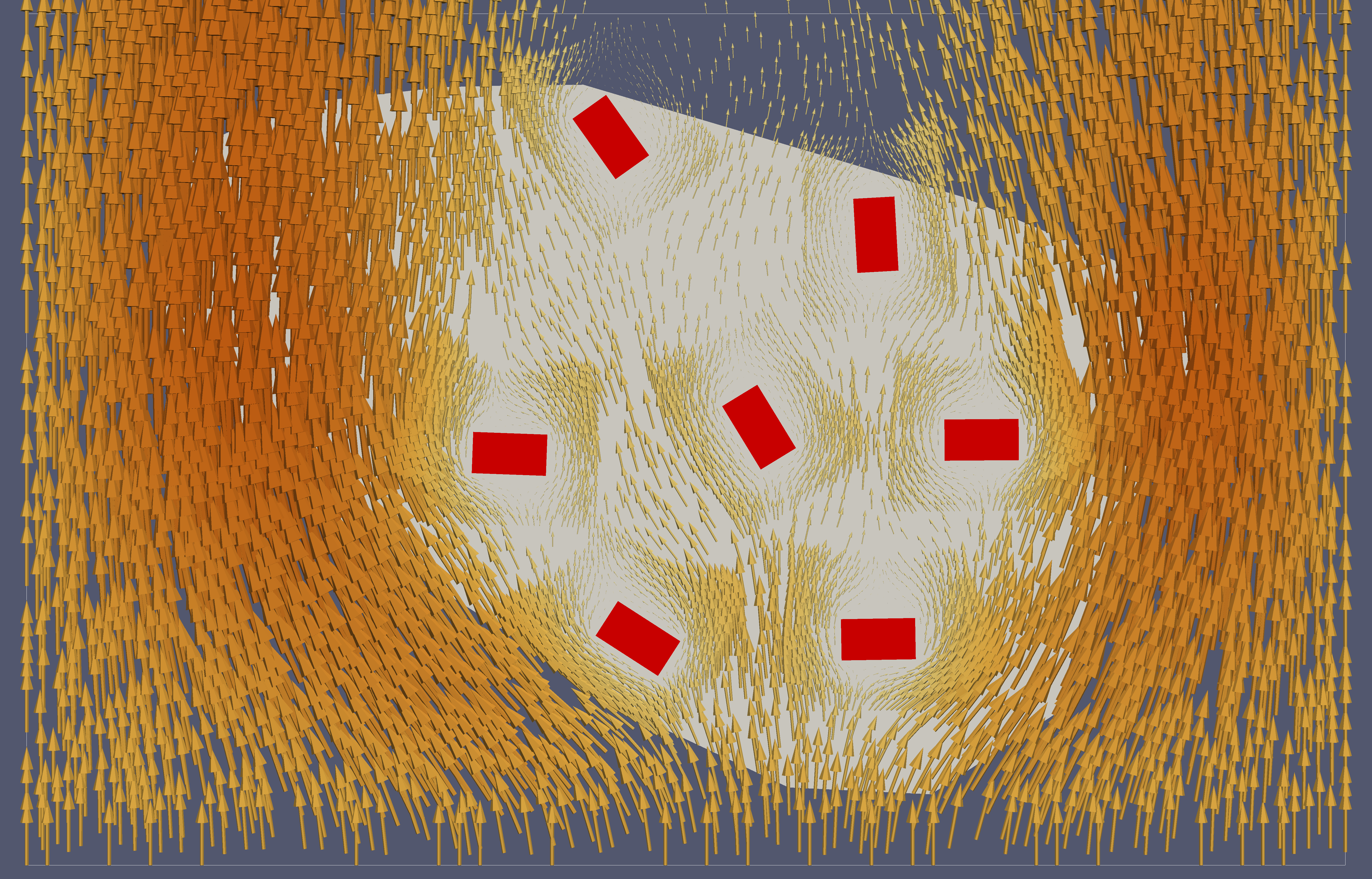}
\includegraphics[width=0.327\linewidth,keepaspectratio]{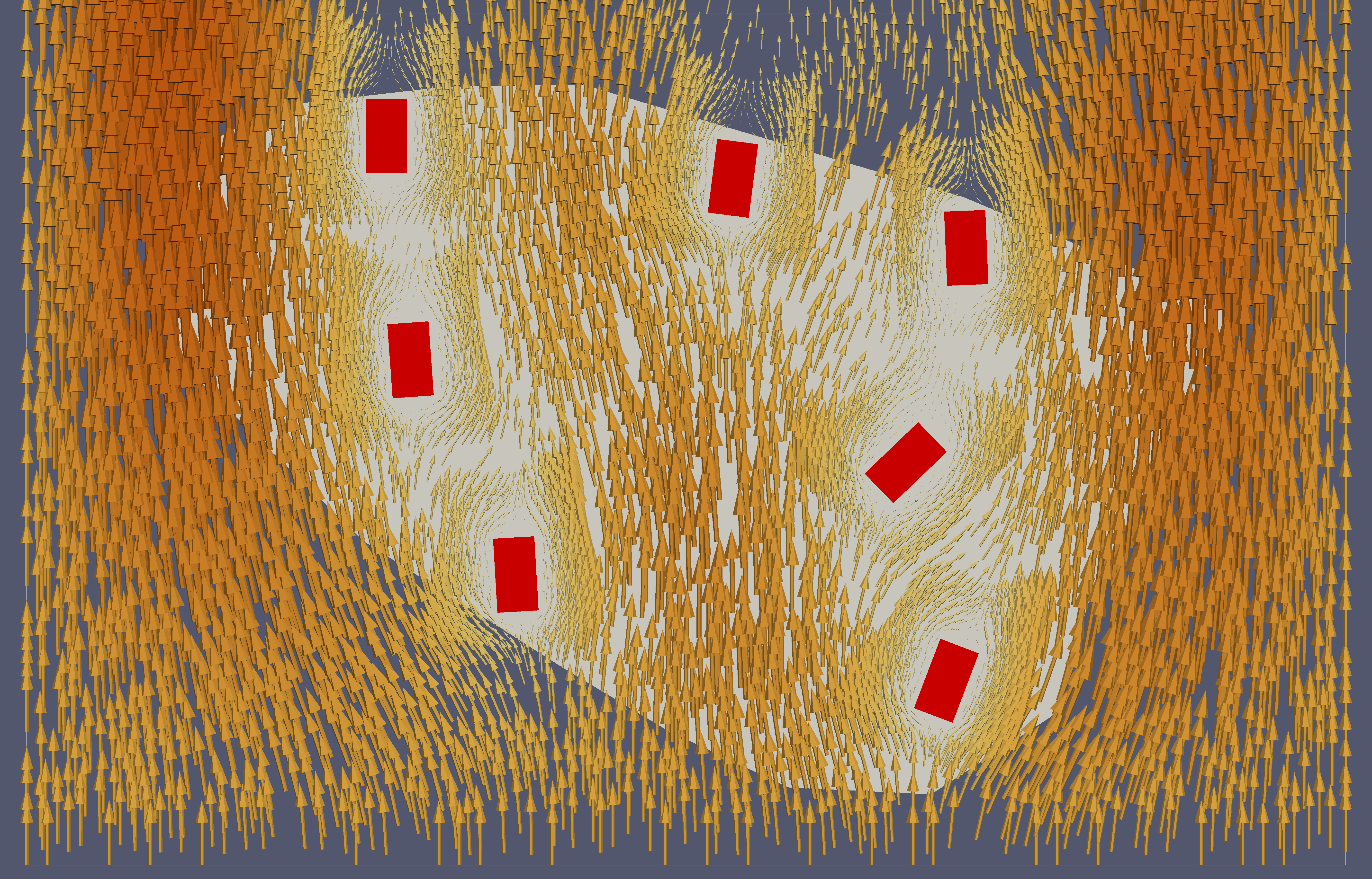}
\includegraphics[width=0.327\linewidth,keepaspectratio]{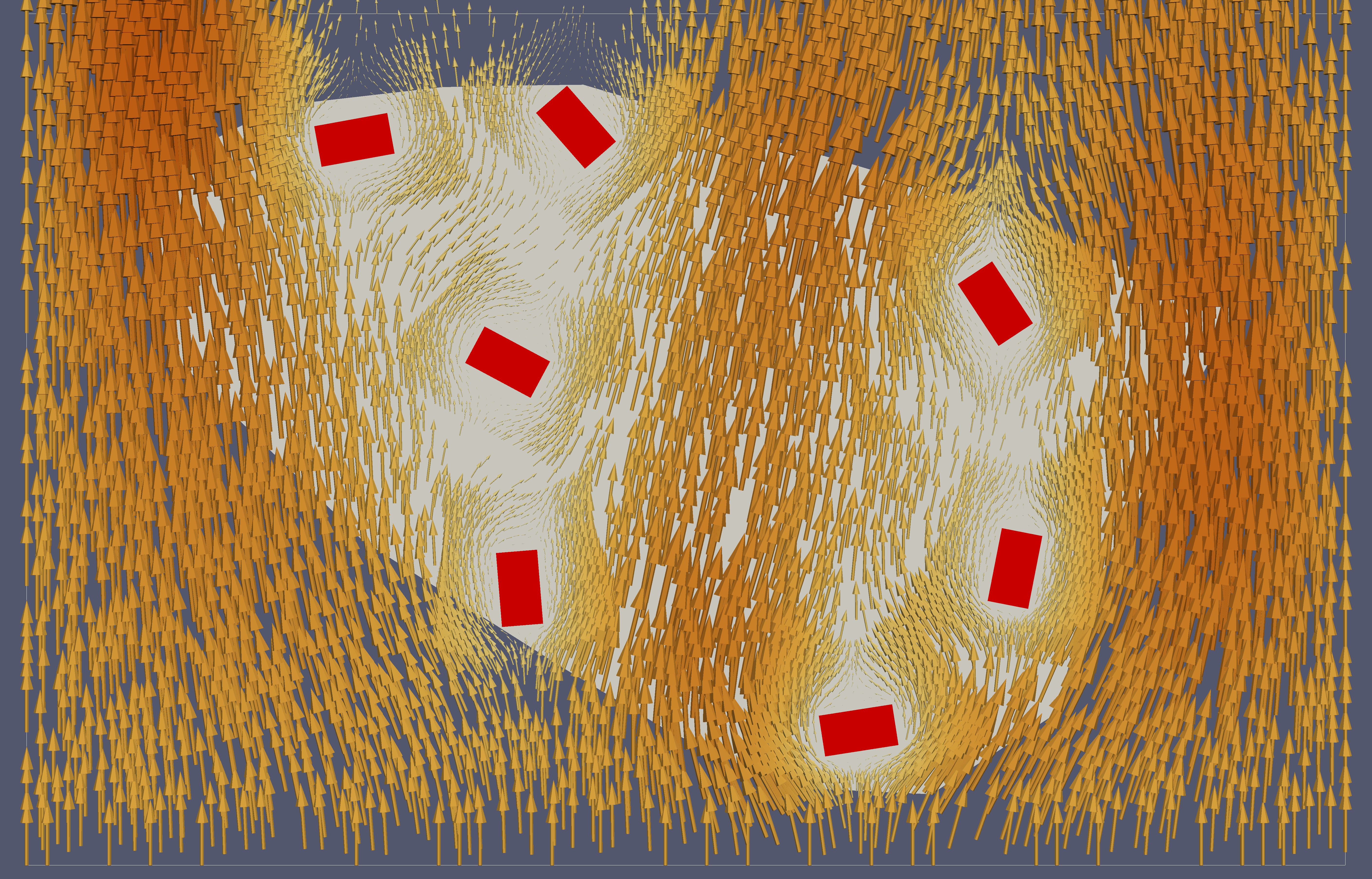}
\includegraphics[width=0.327\linewidth,keepaspectratio]{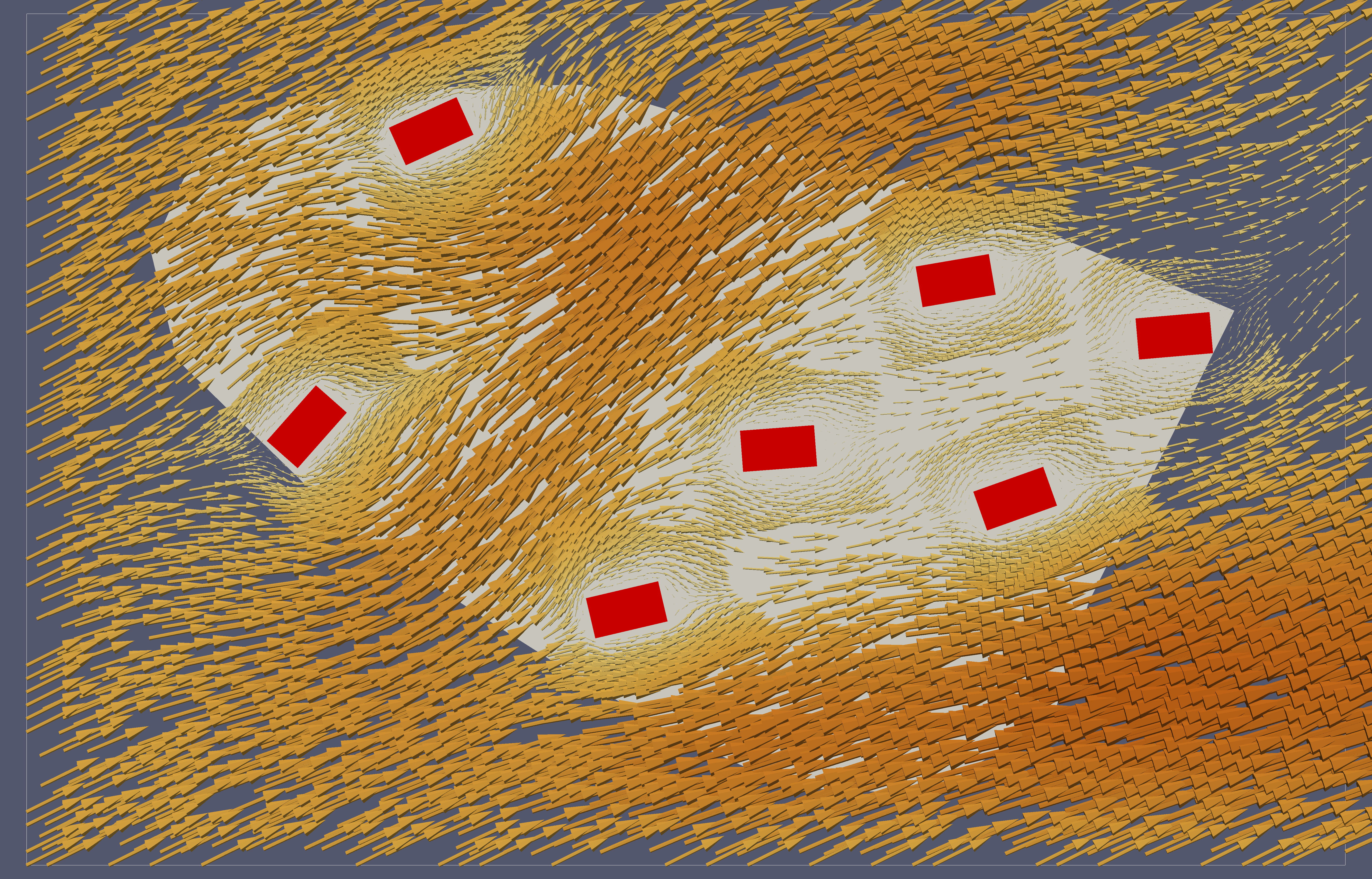}
\includegraphics[width=0.327\linewidth,keepaspectratio]{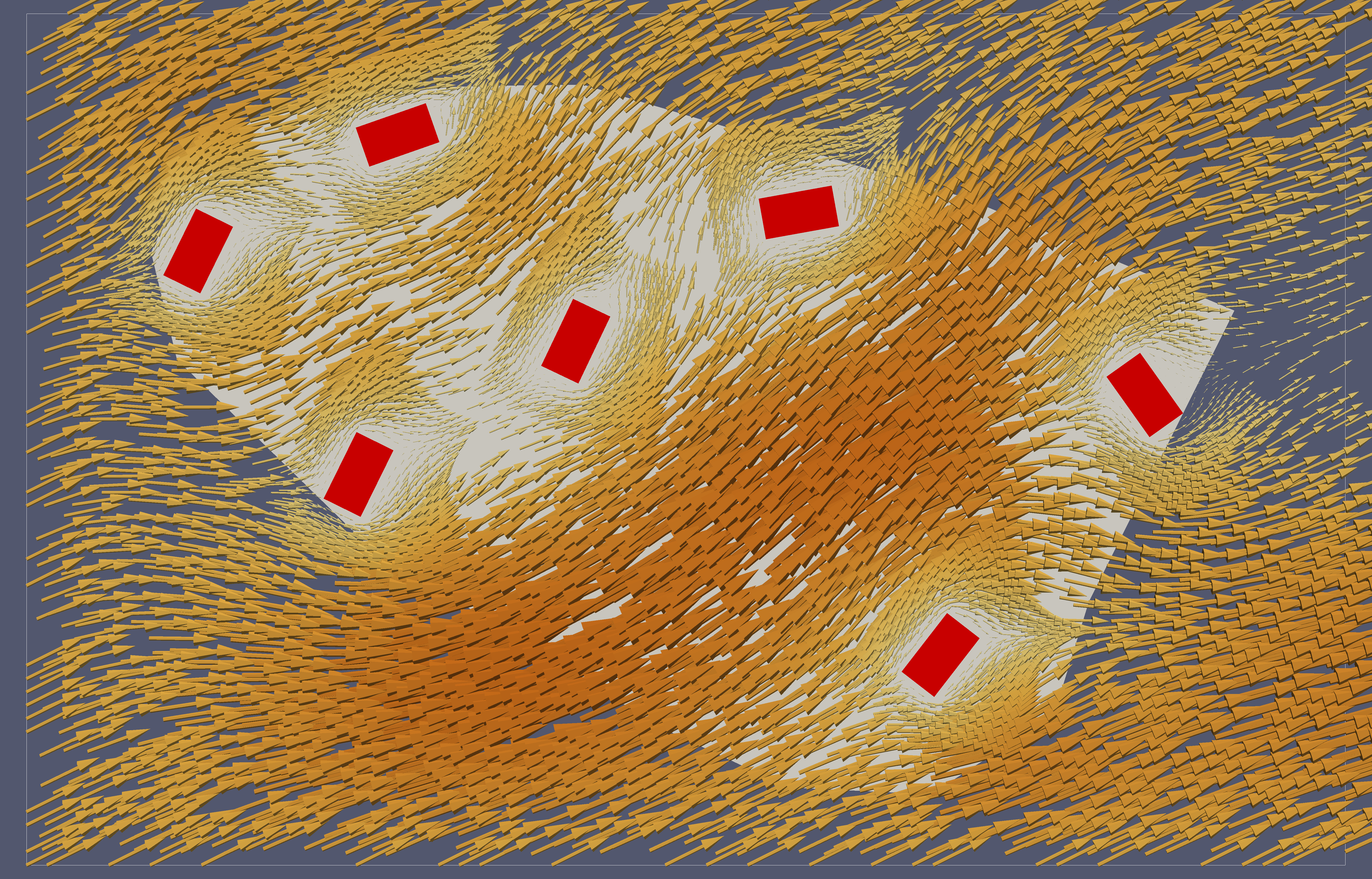}
\includegraphics[width=0.327\linewidth,keepaspectratio]{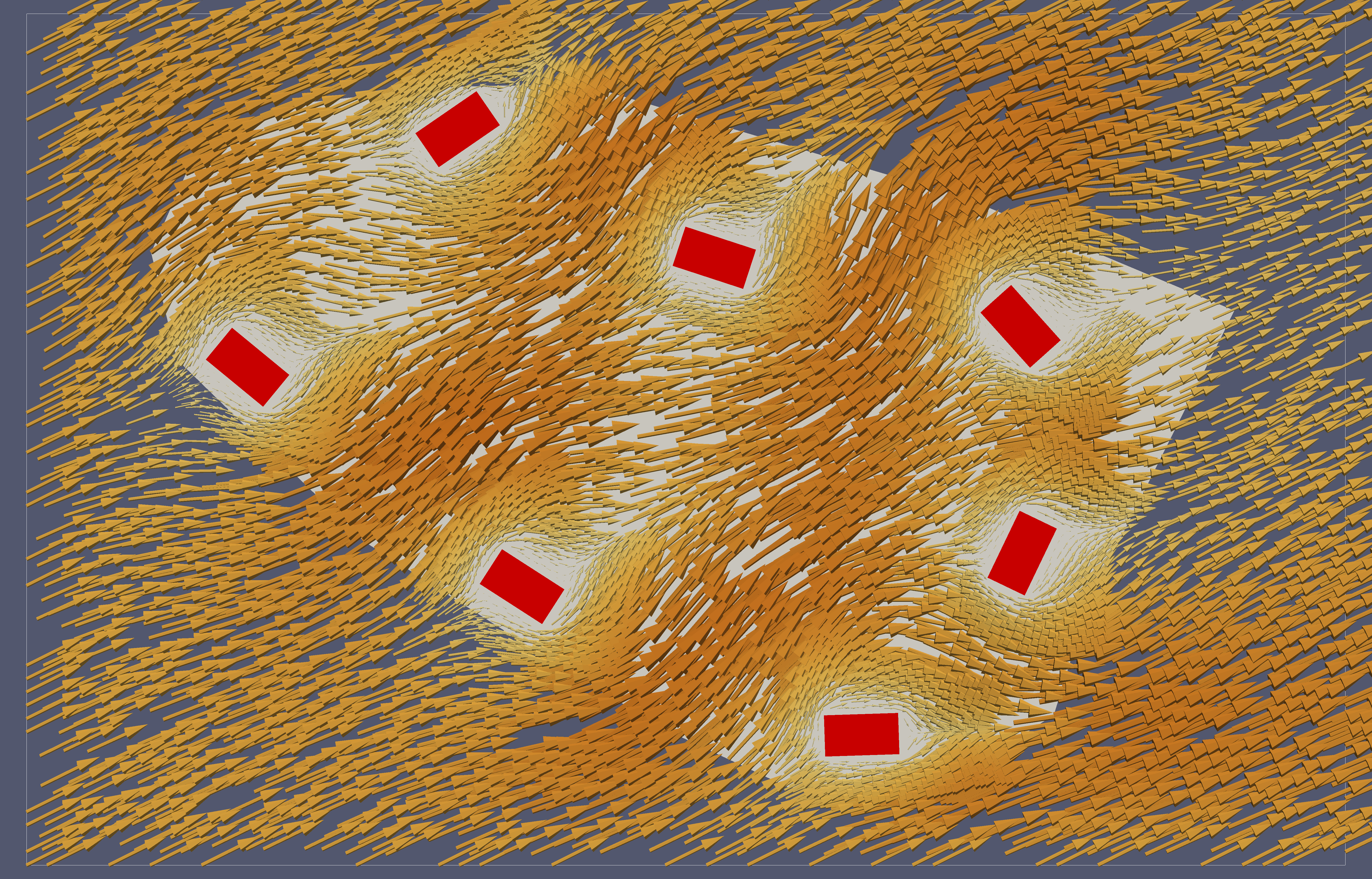}
\caption{Layouts from pure wind optimization. From left to right, the initial layout was hexagonal, principal line, and coastal, respectively. From top to bottom, the wind direction is horizontal, vertical, and diagonal, respectively.}
\label{fig2Dopt_purewind_specified}
\end{figure}

\noindent Layouts from pure wind optimization in 2D are shown in Figure \ref{fig2Dopt_purewind_specified}. The velocity $\mbu_h$ from (\ref{femstokes}) is visualized with glyphs, colored and scaled by magnitude.

\begin{figure}[H]
\centering
\includegraphics[width=0.327\linewidth,keepaspectratio]{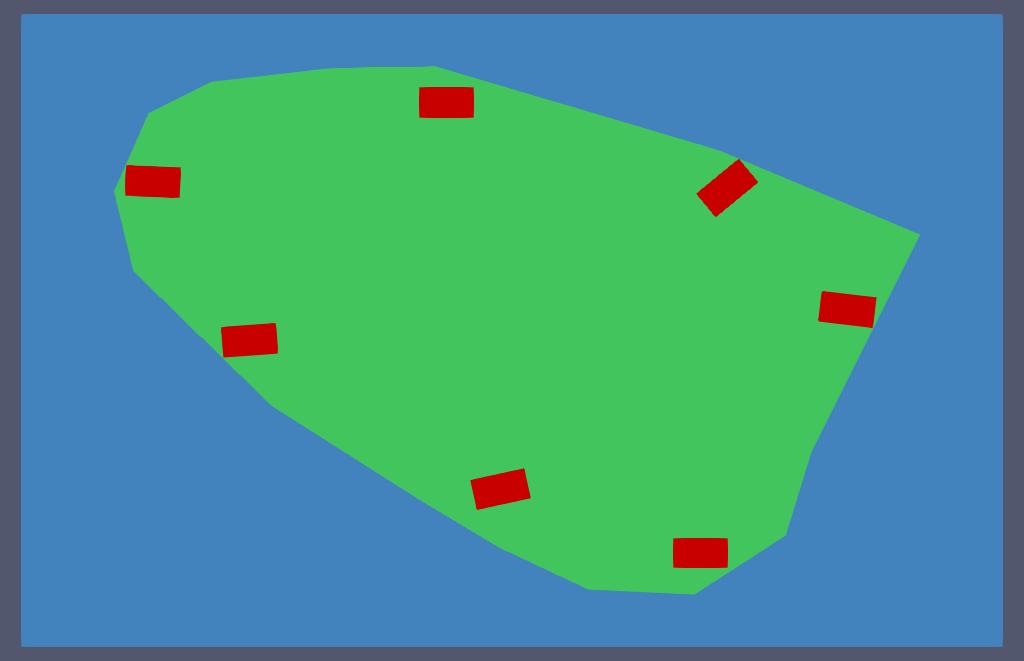}
\includegraphics[width=0.327\linewidth,keepaspectratio]{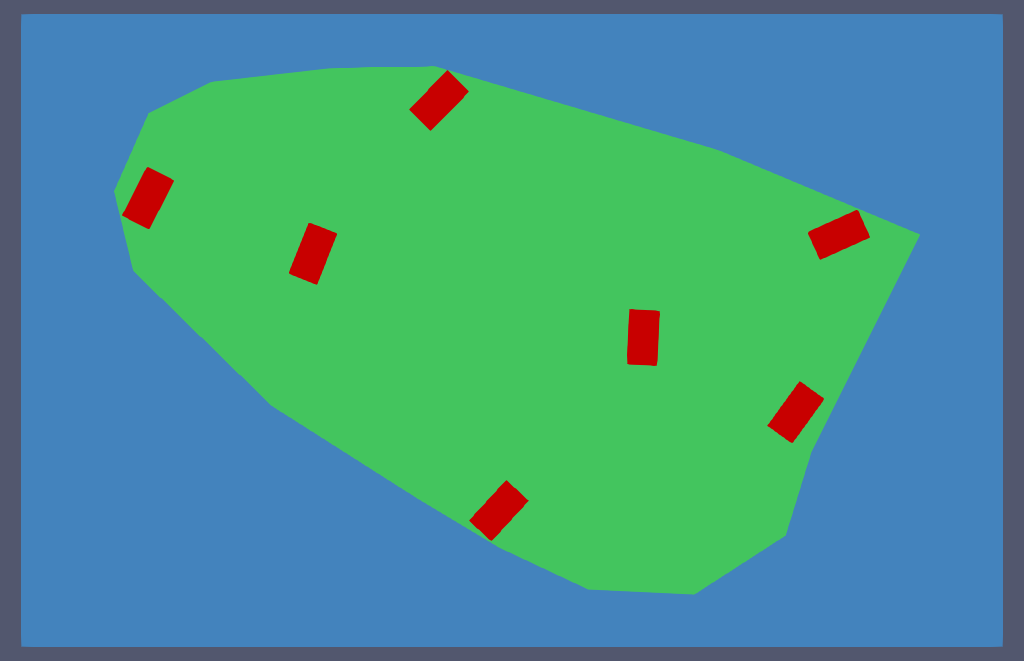}
\includegraphics[width=0.327\linewidth,keepaspectratio]{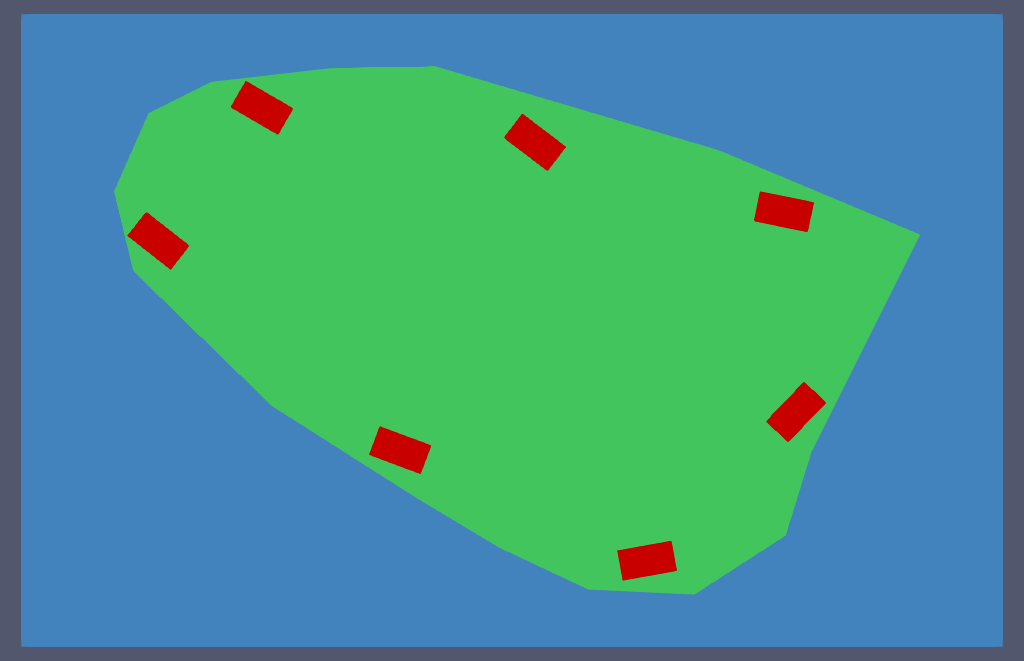}
\caption{Layouts from pure view optimization. From left to right, the initial layout was hexagonal, principal line, and coastal, respectively}
\label{fig2Dopt_pureview_specified}
\end{figure}

Simulation results from pure view optimization in 2D are shown
in Table \ref{tab_opt_pureview_specific} and
Figure \ref{fig2Dopt_pureview_specified}.
\begin{table}[H]
\begin{center}
\begin{tabular}{l | n{1}{5}n{1}{5}n{1}{5}}
    \multicolumn{4}{c}{Pure view optimization} \\ \hline
    Layout & {Hexagonal} & {Principal line} & {Coastal} \\ \hline
    $f_V$ & 0.151327762923 {(\numprint{0.306355916815})} & 0.161918940771 {(\numprint{0.188018240208})} & 0.14858822218 {(\numprint{0.167379980854})} \\
     nfev & {5142} & {6582} & {5040}
\end{tabular}
\caption{Optimized values of the collective measure of view, $f_V$, from pure view optimizations for three initial layouts. Starting values of $f_V$ are shown in parentheses. Number of evaluations of the objective function during optimization, nfev, are shown below.}
\label{tab_opt_pureview_specific}
\end{center}
\end{table}

\begin{figure}[H]
\centering
\includegraphics[width=0.188\linewidth,keepaspectratio]{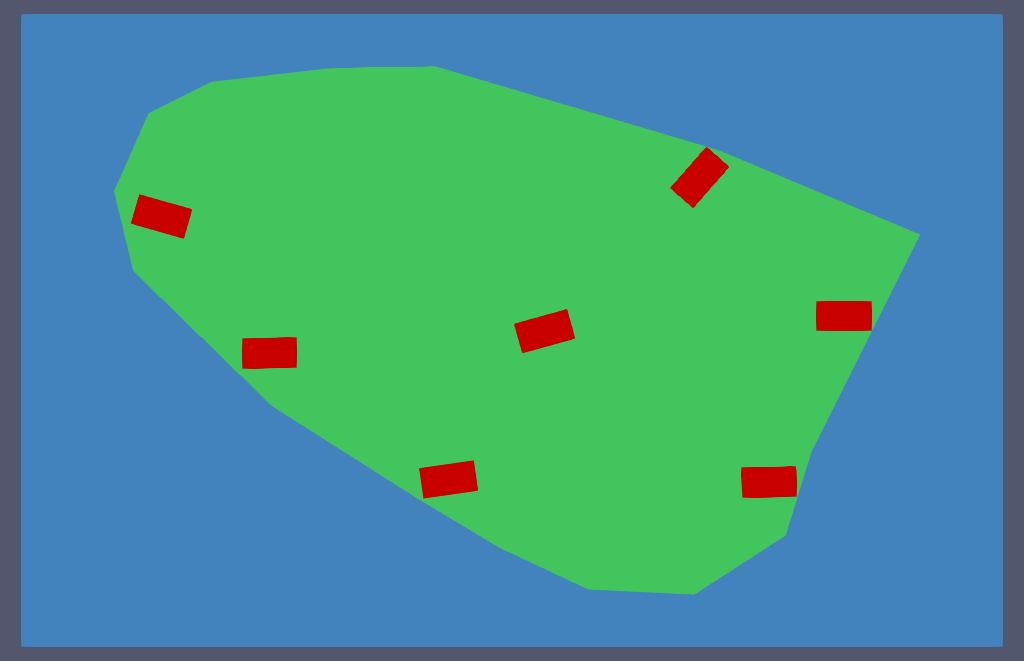}
\includegraphics[width=0.188\linewidth,keepaspectratio]{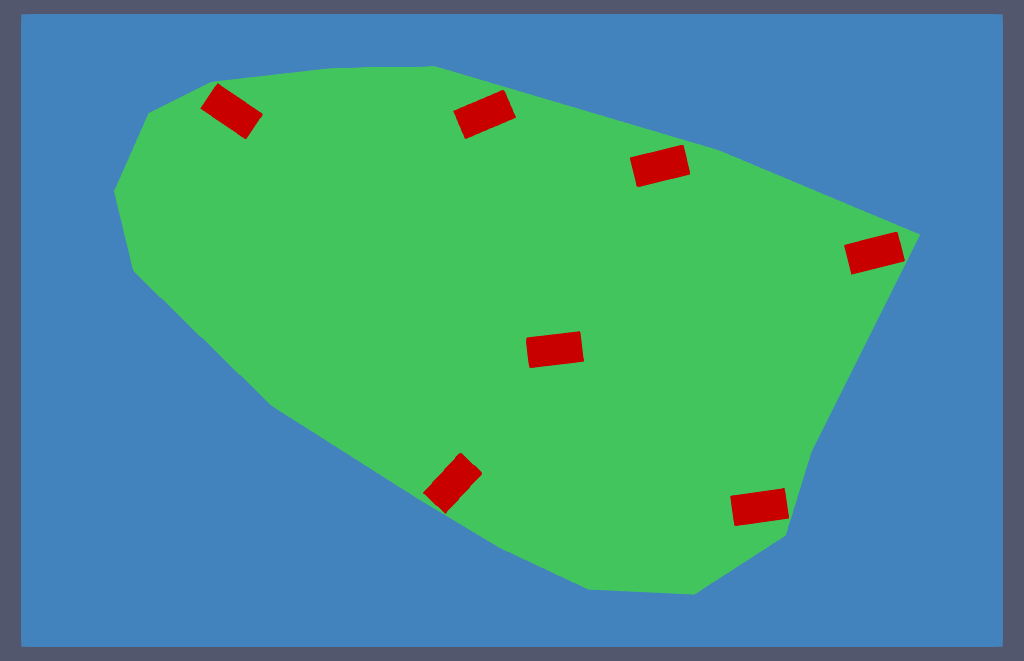}
\includegraphics[width=0.188\linewidth,keepaspectratio]{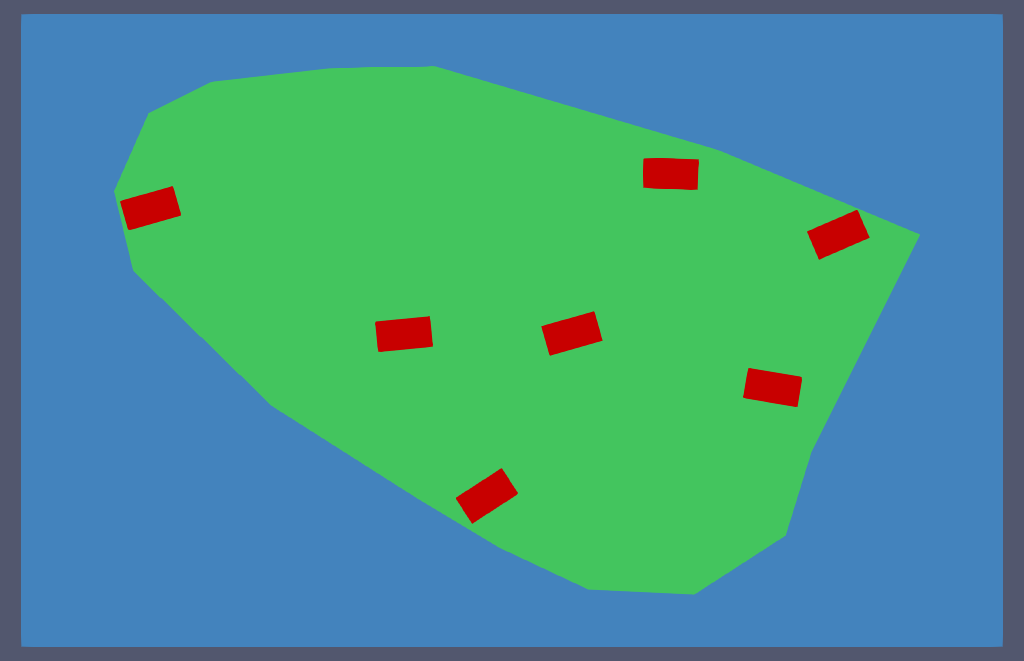}
\includegraphics[width=0.188\linewidth,keepaspectratio]{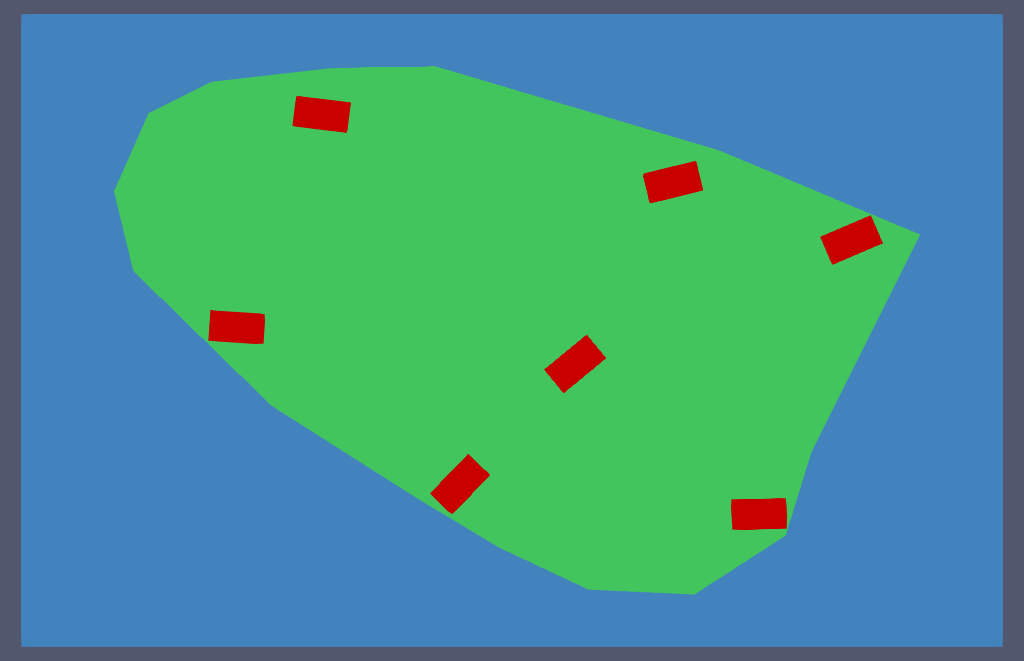}
\includegraphics[width=0.188\linewidth,keepaspectratio]{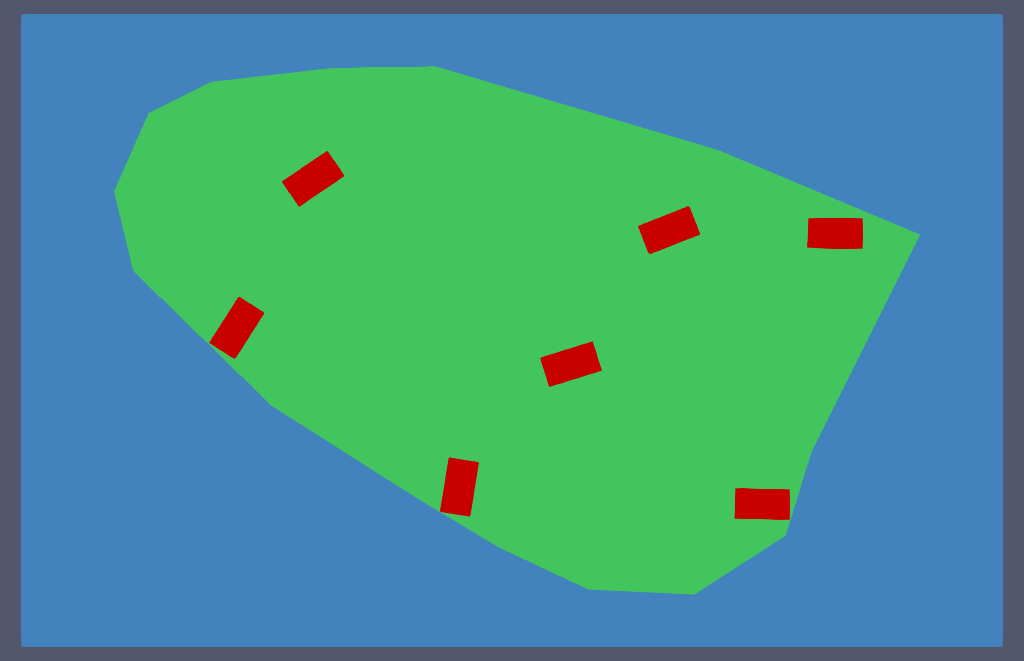}
\includegraphics[width=0.188\linewidth,keepaspectratio]{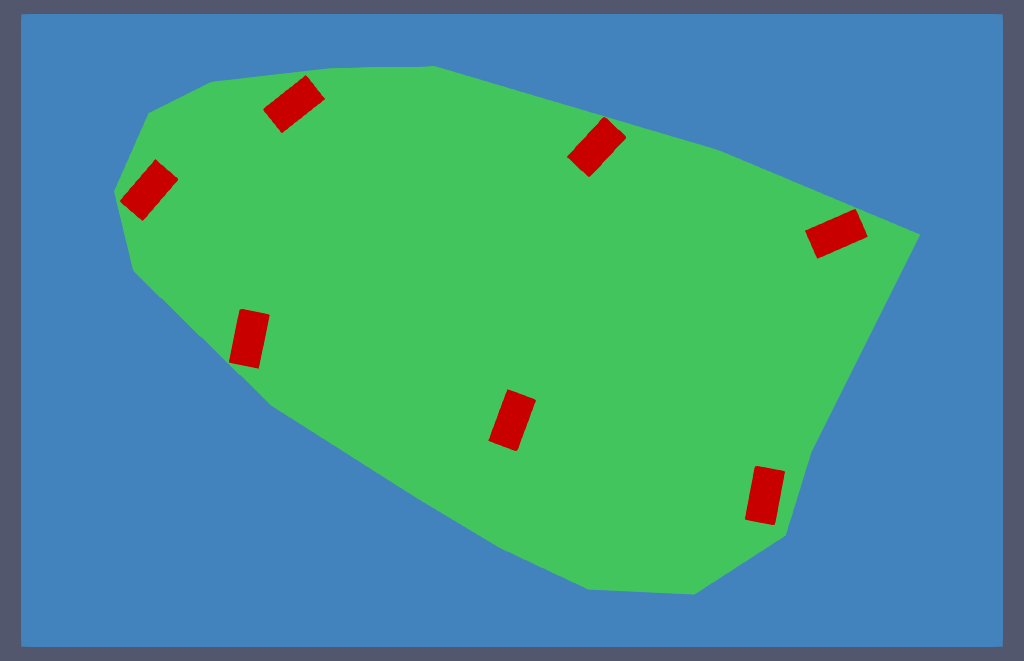}
\includegraphics[width=0.188\linewidth,keepaspectratio]{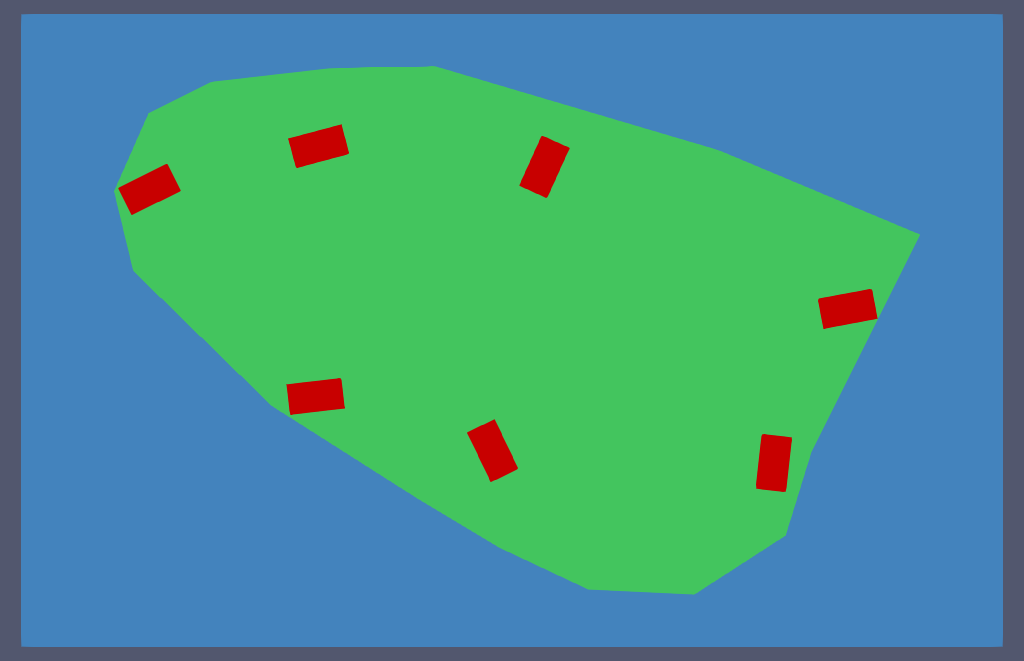}
\includegraphics[width=0.188\linewidth,keepaspectratio]{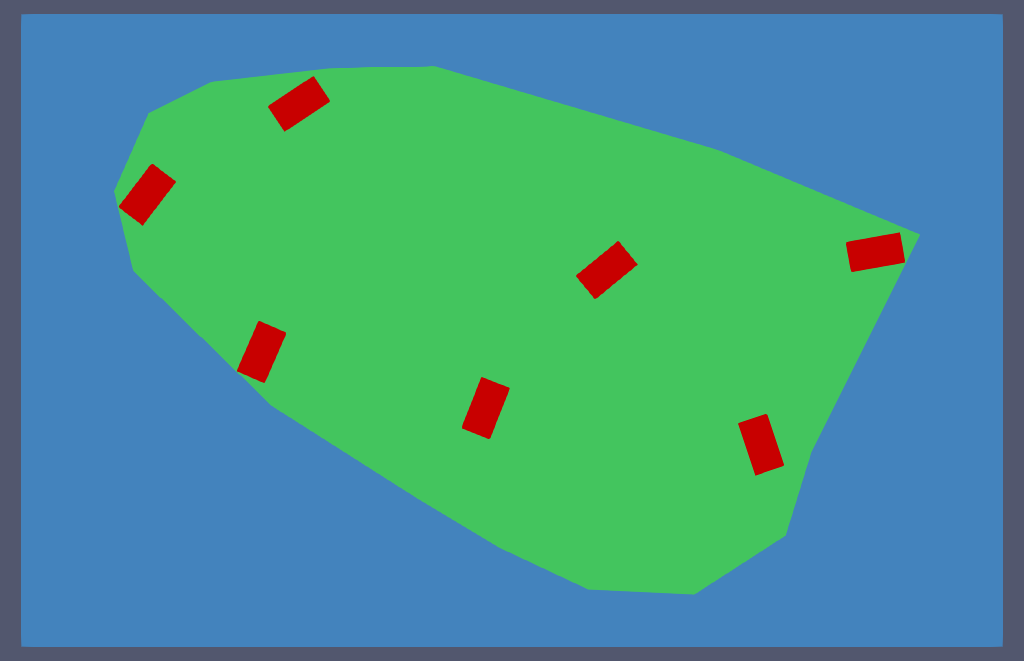}
\includegraphics[width=0.188\linewidth,keepaspectratio]{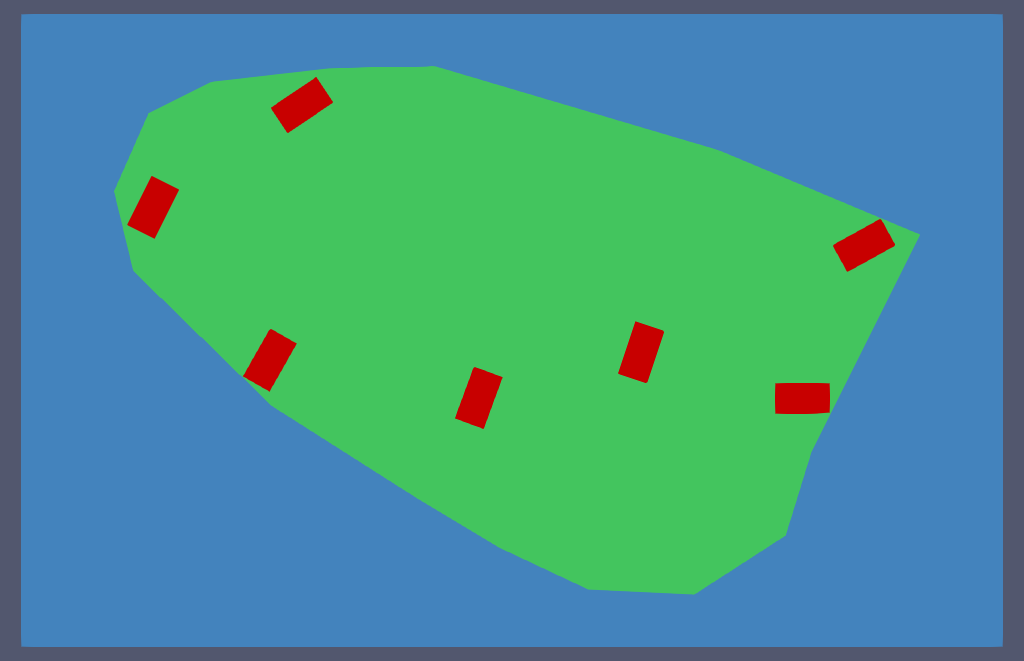}
\includegraphics[width=0.188\linewidth,keepaspectratio]{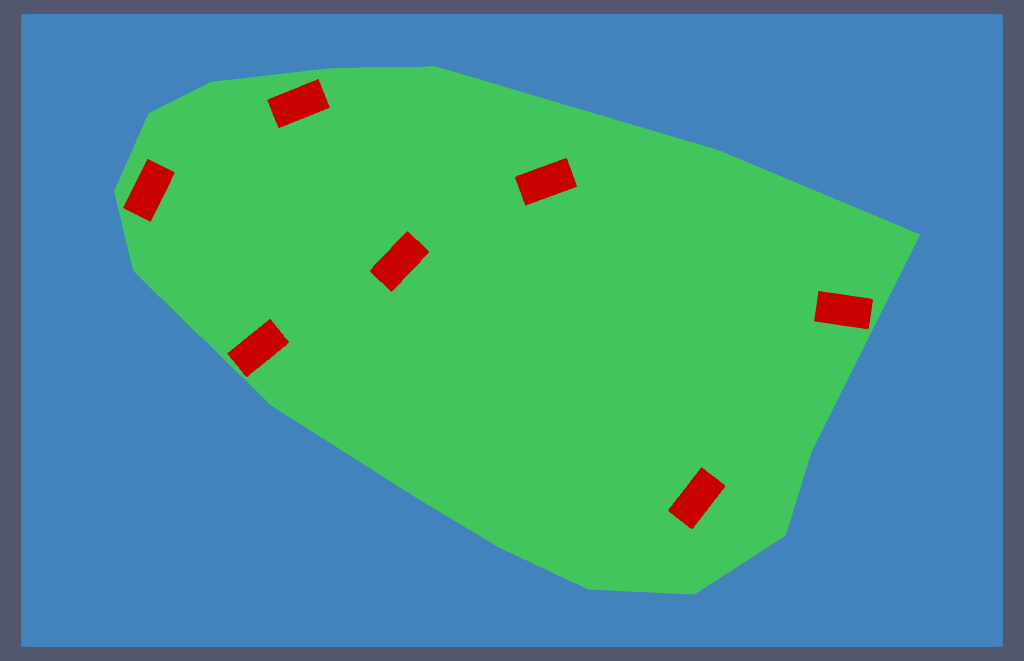}
\includegraphics[width=0.188\linewidth,keepaspectratio]{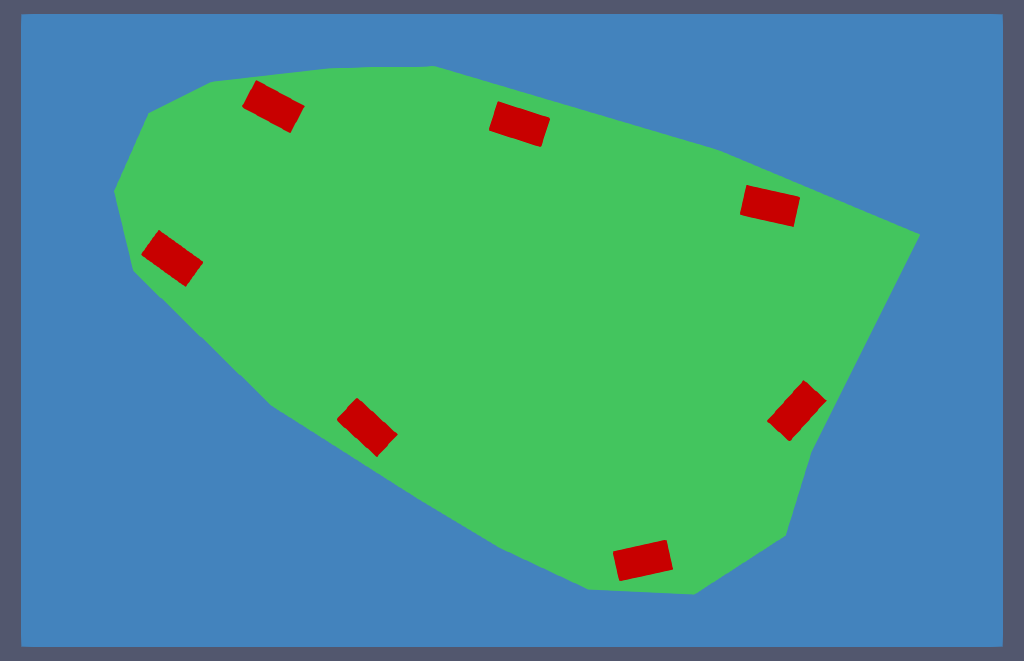}
\includegraphics[width=0.188\linewidth,keepaspectratio]{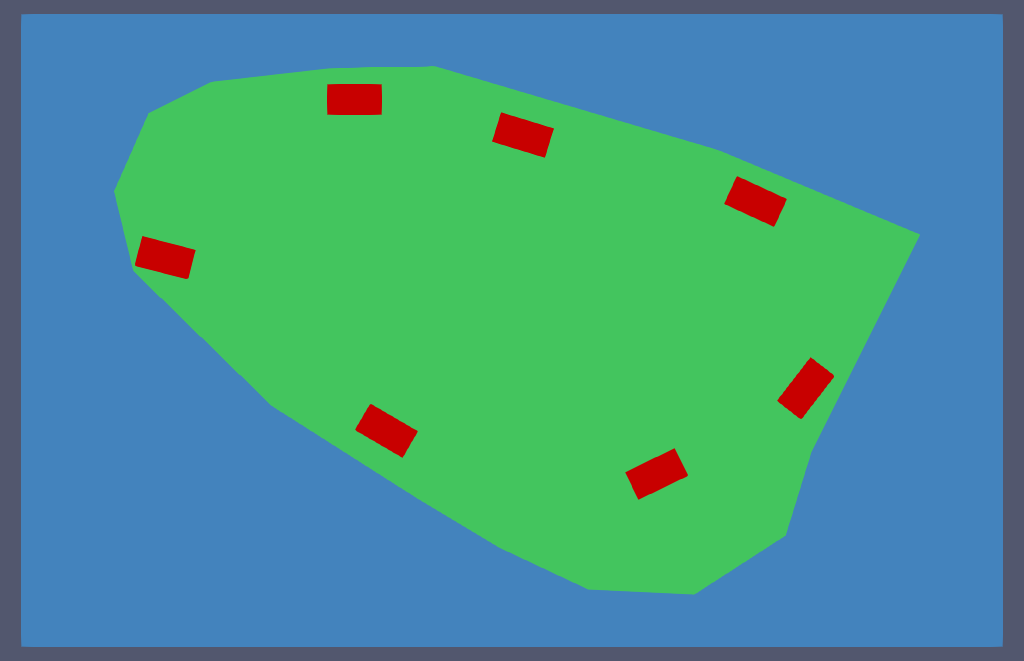}
\includegraphics[width=0.188\linewidth,keepaspectratio]{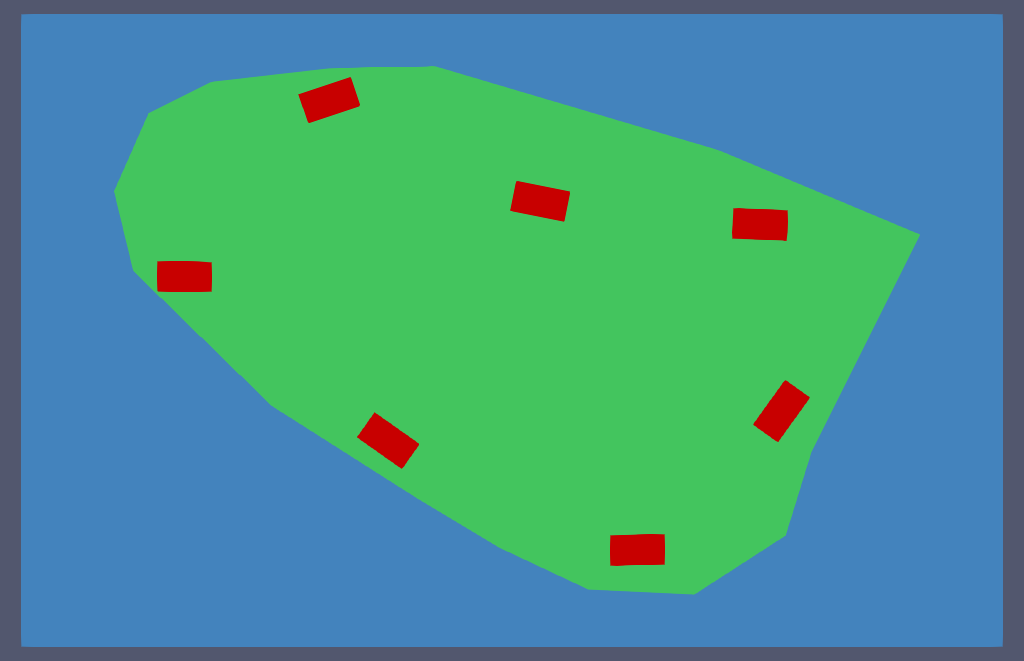}
\includegraphics[width=0.188\linewidth,keepaspectratio]{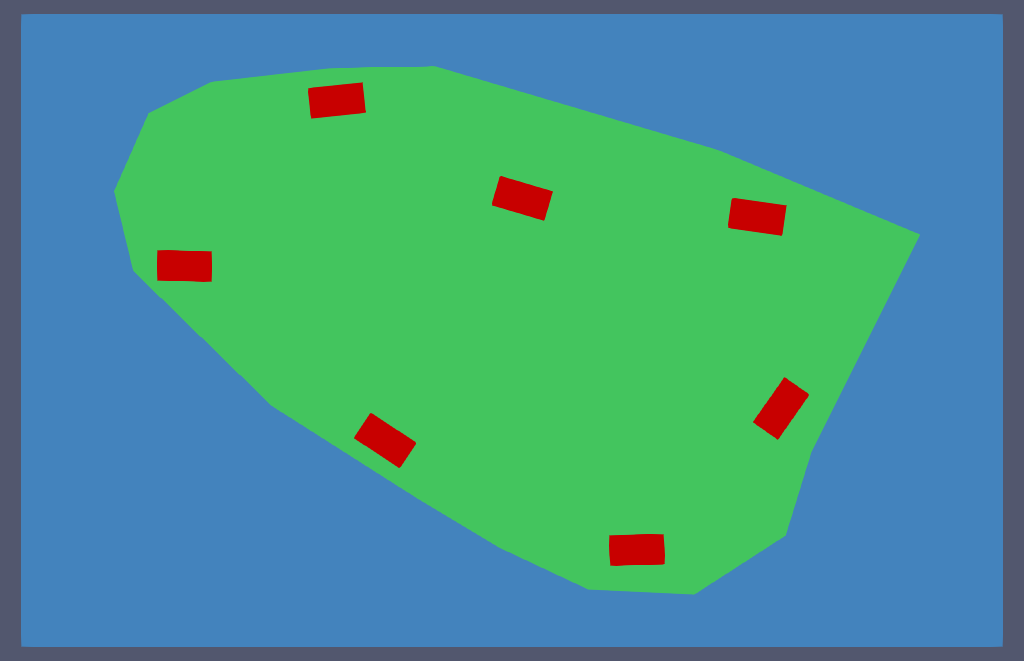}
\includegraphics[width=0.188\linewidth,keepaspectratio]{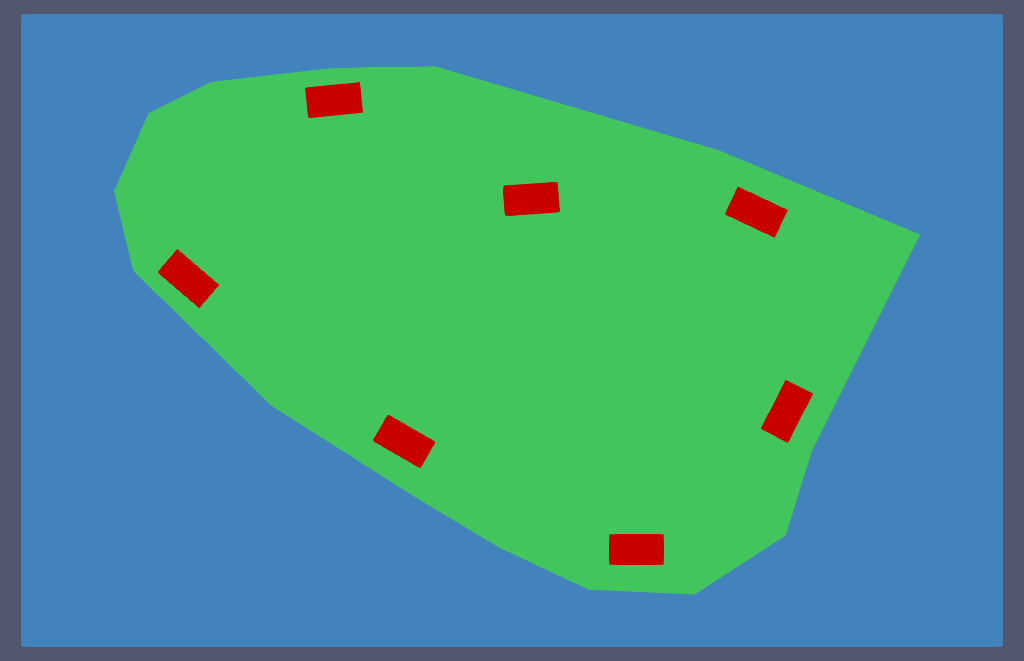}
\caption{Layouts from mixed optimization. From left to right, the wind weight $\alpha = 0.1, 0.3, 0.5, 0.7, 0.9$, respectively. From top to bottom, the initial layout was hexagonal, principal line, and coastal, respectively.}
\label{fig2Dopt_mixed}
\end{figure}

For mixed optimization, only the diagonal inflow wind direction and five values for $\alpha$ are used. Numerical results from mixed optimization in 2D are shown in Table \ref{tab_opt_mixed_specific}.
\begin{table}[H]
\begin{center}
\begin{tabular}{l l | n{1}{5}n{1}{5}n{1}{5}n{1}{5}n{1}{5}}
    \multicolumn{7}{c}{Mixed optimization} \\ \hline
    {Layout} $\setminus$ {$\alpha$} & & {0.1} & {0.3} & {0.5} & {0.7} & {0.9} \\ \hline
                      & $f_W$ {(\numprint{1.586989})} & 1.320726 & 1.192567 & 1.273486 & 1.191630 & 1.183737 \\
    Hexagonal & $f_V$ {(\numprint{0.306355916815})} & 0.164689006213 & 0.184224735891 & 0.20109913526 & 0.180997530995 & 0.233967170151 \\
                      & nfev & {2984} & {5355} & {3001} & {5501} & {3552} \\ \hline
                        & $f_W$ {(\numprint{2.142663})} & 1.263355 & 1.206528 & 1.164209 & 1.158456 & 1.155859 \\
    Principal line & $f_V$ {(\numprint{0.188018240208})} & 0.158388805283 & 0.168116012109 & 0.175665105529 & 0.18978599024 & 0.196924793994 \\
                        & nfev & {3855} & {5282} & {3861} & {6317} & {5248} \\ \hline
                 & $f_W$ {(\numprint{1.391669})} & 1.255290 & 1.224032 & 1.228522 & 1.252896 & 1.270183 \\
    Coastal & $f_V$ {(\numprint{0.167379980854})} & 0.152930122415 & 0.165450556178 & 0.172928328159 & 0.17256385223 & 0.192554202624 \\
                 & nfev & {2480} & {4177} & {3718} & {3143} & {3803} \\
\end{tabular}
\caption{Optimized values of the collective measures of wind, $f_W$, and view, $f_V$, from mixed optimization. Starting values of $f_W$ and $f_V$ are shown in parentheses. Number of evaluations of the objective function during optimization, nfev, are shown below.}
\label{tab_opt_mixed_specific}
\end{center}
\end{table}
\noindent Layouts from mixed optimization in 2D are shown in Figure \ref{fig2Dopt_mixed}.

\section{Discussion}
\label{sec:discussion}

Here we comment on methods and results, starting
with the wind model. Figure \ref{fig2Dflow} and
Figure \ref{fig2Dopt_purewind_specified} show the smooth
transition of $\mbu_h$ between meshes. This indicates that the multi-mesh finite element method produces a solution that transitions seamlessly between meshes. In particular, we
do not want the boundary between the
meshes to show up as a seam in the solution. The glyphs used
to visualize $\mbu_h$ in Figure \ref{fig2Dflow} and Figure
\ref{fig2Dopt_purewind_specified} obviously reveal the existence
and extent of overlapping house meshes. But this is because the overlapping
house meshes have a finer mesh size than the background air mesh, see
Figure \ref{figmeshoverlap}, and that the glyphs are placed uniformly over
the nodal points in the meshes, resulting in a higher glyph density
on the house meshes.


Concerning the 3D view model, Figure \ref{fig3Dview}
shows some views from the selected location.
The top left frame shows a fair view with a
lot of sea and sky but also three houses and a part of the island.
Therefore the view valuation becomes $V=0.79$. To get a
better view there should be as much sea or sky as possible, fewer
houses, and less ground. This can be seen in the top right frame,
where $V = 0.90$. Examples of views from the other
end of the view scale can be found in the bottom frame.
In the bottom left frame one almost only sees a
neighboring house, resulting  in a very low view value.
In \cite{Bourassa2004} Table 1, previous studies on the
impact of view on real estate prices are listed.
Every study classifies the view differently. Summarizing
these studies, view from a waterfront increases the value
of a house by $127\%-147\%$, full water view by $32\%-68\%$,
and simple/partial water view by $6\%-10\%$. It is hard to directly
compare these numbers to our view computations. But comparing the
full waterfront view in the top right frame of Figure \ref{fig3Dview} with the partial sea
view in the bottom left frame, the waterfront view ($V=0.90$) is $125\%$
higher than the partial sea view ($V=0.40$).


For the 2D view model, Figure \ref{fig2Dview}, shows that $f_V$ is larger
when the houses are lumped together (hexagonal with $f_V = \numprint{0.306355916815}$),
than when spread out (coastal $f_V = \numprint{0.167379980854}$).
This is what to be expected from the construction of the view measure,
since seeing more house should result in poorer view, and seeing
less house in a better. When the houses are placed closely, but on a line,
$f_V$ is also relatively low (principal line with $f_V = \numprint{0.188018240208}$).
This is because the view from a house consists of at most two other houses,
behind which the rest of the houses are hidden.

The optimization results show that there are multiple local minima to
our optimization problem. For pure wind optimization, Table
\ref{tab_opt_purewind_specific} shows that $f_W$ decreases for all
nine cases. It also shows that the starting values of $f_W$ are
significantly higher ($f_W > 2$) for principal line layout with vertical
and diagonal wind. This is probably because a lot of wind is forced in
between the closely placed houses, resulting in high surrounding
wind speeds, which also is to be expected. From Table \ref{tab_opt_purewind_specific},
the optimized values of $f_W$ seem to get smaller when more
evaluations of the objective function are used in the optimizations.
The optimizations that resulted in the three smallest values all used
more than 10000 evaluations. Figure \ref{fig2Dopt_purewind_specified}
shows that it can be beneficial for the houses to align themselves with the wind and to
lie in the wake of other houses, similar to how birds fly in V formation.
This is especially evident for the middle layout on the middle row of Figure
\ref{fig2Dopt_purewind_specified} ($f_W = \numprint{0.863266}$),
and for the left layout on the bottom row of Figure \ref{fig2Dopt_purewind_specified}
($f_W = \numprint{0.992282}$).


For pure view optimization, Table
\ref{tab_opt_pureview_specific} shows that $f_V$ decreases to similar
values for all three cases. The starting value of $f_V$ for hexagonal initial
layout is significantly larger than its optimized value. This is in contrast to
the other initial layouts, for which starting and optmized values are similar.
This suggest that principal line and coastal already are good layouts
with respect to view. Comparing Figure \ref{fig2Dopt_pureview_specified} with
Figure \ref{fig2Dview}, this seems to be true since the left case
goes from hexagonal to a coastal layout, and the right case shows
that the coastal layout barely changes. The middle case seems
to be a compromize between principal line and coastal layout,
with some houses at the coast and others along the principal line
of the island.

For mixed optimization, we would expect $f_W$ to decrease
when $\alpha$ increases. Table \ref{tab_opt_mixed_specific}
shows that for hexagonal, the values are generally decreasing;
for principal line, the values are strictly decreasing, and
for coastal the values remain quite constant (actually increasing somewhat
for $\alpha > 0.3)$. In the bottom row of Figure \ref{fig2Dopt_mixed} the layouts
do not change much, suggesting that the optimization process
gets stuck in a local minimum of the optimization landscape
when using the coastal layout, leading to quite constant values of $f_W$.
Considering $f_V$ for mixed optimization, we would expect it to increase,
when $\alpha$ increases. Table \ref{tab_opt_mixed_specific} shows that
for hexagonal and coastal, the values are generally increasing
(almost strictly for coastal), and for principal line the values are strictly increasing.
For the layouts from mixed optimization, we would expect them to have
characteristics of pure view layouts, i.e., more spread out, for small $\alpha$.
As $\alpha$ increases, we would expect to see less of these and
more characteristics of pure wind layouts, i.e., alignment with wind and lying
in wakes. For hexagonal (top row of Figure \ref{fig2Dopt_mixed}), characteristics of
pure wind layouts seem to be present in all cases, but to be more apparent
for larger $\alpha$. For principal line (middle row of Figure \ref{fig2Dopt_mixed}), the
behavior of the layouts seems to be as expected when $\alpha$ increases, i.e., going from a spread out
layout to one where houses lie in wakes of other houses and
are aligned with the diagonal wind. As already mentioned,
the behavior observed for coastal (bottom row of Figure \ref{fig2Dopt_mixed}), suggests the entrapment
in a local minimum valley.

\section{Conclusions and outlook}
\label{sec:conclusions}

We have presented a generic framework for evaluating and optimizing
settlement layouts with respect to both wind and view.
The framework allows multiple configurations to be
computed and evaluated with relative ease thanks to the application
of multi-mesh finite element methods. The current
proof-of-concept implementation has several limitations that will be
addressed in future work. These limitations and extensions fall into
three different categories: modelling, multi-mesh software, and ease of
use.

Concerning modelling, the use of Stokes equations in 2D to
model airflow between buildings has two important limitations:
Stokes flow is not physically appropriate to model airflow, and
real world buildings are 3D. Future work would therefore extend the
wind model to more advanced flow equations in 3D.  For the computation of
view, a more extensive comparison with real world data, in order to calibrate
functions and weights, could be a topic for further research, perhaps in a
similar fashion to \cite{Fisher-Gewirtzman2018}.
Although the current basic optimization seems to work fairly well,
there is a lot of potential for future studies and improvements. We mainly
think of the definition of the objective function, and the choice of algorithm
for solving the optimization problem.

Regarding multi-mesh software, we noticed during this study that some
particular configurations of buildings in 3D resulted in numerical
instabilities. This is likely due to bugs or untreated corner cases in
the computational geometric framework of the FEniCS multi-mesh
implementation. This issue is also the focus of ongoing work.

To be a useful tool in an iterative architectural process, the
framework must not only be efficient and robust. It must also be easy
to use. Future work will also consider the creation of a user-friendly
interface. In particular, it would be highly relevant to consider the
creation of virtual or mixed reality interfaces to our framework.

\section*{Acknowledgements}

This work was supported by the Swedish Research Council
(Grant No. 2014-6093) and by the profile area \emph{Virtual Cities} as
part of \emph{Building Futures}, an Area of Advance at Chalmers.
We are also grateful to CREAM Architects in Gothenburg for ideas,
feedback and inspiration during this project; and Professor
Michael Patriksson, Chalmers University of Technology and University
of Gothenburg, for consultation and guidance regarding optimization.



\bibliographystyle{elsarticle-num}
\bibliography{bibliography}

\end{document}